\documentclass[useAMS,usenatbib,fleqn]{mn2e}
\usepackage{graphicx}
\usepackage{epstopdf}
\usepackage{amsmath}
\usepackage{amsfonts}
\usepackage{amssymb}
\usepackage[draft]{hyperref}
\usepackage[usenames, dvipsnames]{color}

\def\ul{\char`_}

\newcommand{\comm}[1]{\normalfont{#1}}

\title[The LMC outer stellar populations and geometry]{The LMC geometry and outer
stellar populations \\ from early DES data}

\author[E. Balbinot et al.]{Eduardo Balbinot$^{1,2,3}$\thanks{e-mail: e.balbinot@surrey.ac.uk},
                            B. X. Santiago$         ^{2,3}$,
                            L. Girardi$             ^{4}$,
                            A. Pieres$              ^{2,3}$,
                            L. N. da Costa$         ^{3,5}$,
                            \newauthor
                            M. A. G. Maia$          ^{3,5}$,
                            R. A. Gruendl$          ^{6,7}$
                            A. R. Walker$           ^8$,
                            B. Yanny$               ^9$,
                            A. Drlica-Wagner$       ^{9}$,
                            \newauthor
                            A. Benoit-Levy$         ^{10}$,
                            T. M. C. Abbott$        ^{8}$,
                            S. S. Allam$            ^{9,11}$,
                            J. Annis$               ^{9}$,
                            J. P. Bernstein$        ^{12}$,
                            \newauthor
                            R. A. Bernstein$        ^{13}$,
                            E. Bertin$              ^{14}$,
                            D. Brooks$              ^{10}$,
                            E. Buckley-Geer$        ^{9}$,
                            A. Carnero Rosell$      ^{3,5}$,
                            \newauthor
                            C. E. Cunha$            ^{15}$,
                            D. L. DePoy$            ^{16}$,
                            S. Desai$               ^{17,18}$,
                            H. T. Diehl$            ^{9}$,
                            P. Doel$                ^{10}$,
                            J. Estrada$             ^{9}$,
                            \newauthor
                            A. E. Evrard$           ^{19,20,14}$,
                            A. Fausti Neto$         ^{3}$,
                            D. A. Finley$           ^{9}$,
                            B. Flaugher$            ^{9}$,
                            J. A. Frieman$          ^{9,18}$,
                            \newauthor
                            D. Gruen$               ^{17,21}$,
                            K. Honscheid$           ^{22}$,
                            D. James$               ^{8}$,
                            K. Kuehn$               ^{23}$,
                            N. Kuropatkin$          ^{9}$,
                            O. Lahav$               ^{10}$,
                            \newauthor
                            M. March$               ^{24}$,
                            J. L. Marshall$         ^{16}$,
                            C. Miller$              ^{19,20}$,
                            R. Miquel$              ^{25,26}$
                            R. Ogando$              ^{3,5}$,
                            J. Peoples$             ^{9}$,
                            \newauthor
                            A. Plazas$              ^{27}$,
                            V. Scarpine$            ^{9}$,
                            M. Schubnell$           ^{12}$,
                            I. Sevilla-Noarbe$      ^{28}$,
                            R. C. Smith$            ^{8}$,
                            \newauthor
                            M. Soares-Santos$       ^{9}$,
                            E. Suchyta$             ^{22}$,
                            M. E. C. Swanson$       ^{7}$,
                            G. Tarle$               ^{19}$,
                            D. L. Tucker$           ^{9}$,
                            \newauthor
                            R. Wechsler$            ^{29,18}$,
                            J. Zuntz$               ^{30}$ \\
                            \emph{(Affiliations can be found after the references)}
                        }

\begin{document}
\pagerange{\pageref{firstpage}--\pageref{lastpage}}
\maketitle
\label{firstpage}

\begin{abstract}
\comm{The Dark Energy Camera has captured a large set of images as part of
Science Verification (SV) for the Dark Energy Survey.} The \comm{SV footprint}
covers a large portion of the outer Large Magellanic Cloud (LMC), providing
\comm{photometry 1.5 magnitudes} fainter than the main sequence turn-off of the
oldest LMC stellar population.  We derive geometrical and structural parameters
for \comm{various} stellar populations in the LMC disk.  For the distribution
of all LMC stars, we find an inclination of $i=-38.14^{\circ}\pm0.08^{\circ}$
(near side in the North) and a position angle \comm{for} the line of nodes of
$\theta_0=129.51^{\circ}\pm0.17^{\circ}$.  We find that stars younger than
$\sim 4$ Gyr are more centrally concentrated than older stars.  \comm{Fitting}
a projected exponential disk shows that the scale radius of the old populations
is $R_{>4 Gyr}=1.41\pm0.01$ kpc, while the younger population has $R_{<4
Gyr}=0.72\pm0.01$ kpc.  \comm{However,} the spatial distribution of the younger
population deviates significantly from the projected exponential disk model.
\comm{The distribution of old stars suggests a large truncation radius of
$R_{t}=13.5\pm0.8$ kpc.} If this truncation is dominated by the tidal field of
the Galaxy, we find that the LMC is $\simeq 24^{+9}_{-6}$ times less massive
than the encircled \comm{Galactic} mass. \comm{By measuring the Red Clump peak
magnitude and comparing with the best-fit LMC disk model, we find that the
LMC disk is warped and thicker in the outer regions north of the LMC
centre. Our findings may either be interpreted as a warped and flared
disk in the LMC outskirts, or as evidence of a spheroidal halo component.}
\end{abstract}


\section{Introduction}

The Milky Way (MW) satellite system offers a variety of examples of dwarf
galaxies. Most of its members are essentially gas-free and contain mainly old
stars \citep{McConnachie12}. The evolution of these systems is closely
related to the formation of the Galaxy and the process of mass assembly of the
large scale structures in the Universe \citep{Klypin99, Moore99, Stewart08}.
On the other hand, the Large and Small Magellanic Clouds (LMC and SMC,
respectively) are the closest low-mass, gas-rich \citep{Grcevich09} interacting
systems. The main features tracing the interaction history of the
clouds are the HI Magellanic Stream \citep{Mathewson74}, and Bridge
\citep{Hindman63}.  A counterpart of the Stream was found and named the Leading
Arm \citep{Putman98}. The formation of these structures is a subject of great
debate. Recent simulations favour a scenario where the Stream, Bridge, and
Leading Arm are remains from the close interaction between the LMC and SMC
before falling into the MW potential \citep{Besla12, Kallivayalil13}.

The Clouds' star formation history (SFH) also reflects their close
interaction history. It is possible to identify multiple periods of
enhanced star formation that are arguably correlated with close encounter
between the Clouds \citep{Meschin14, Rubele12, Javiel05, Holtzman99}. Evidence
of such events are imprinted in the stellar population which, due to the
proximity of the Magellanic System, is resolved into individual stars with
medium sized ground-based telescopes. Enhanced star formation is also
demonstrated by the extensive star cluster system throughout the Magellanic
Clouds (MC). The clusters in this system span a very broad range in age
and metallicity \citep{Kerber09}. Evidence of an age-gap \citep{Jensen88}
may support a relationship between the formation/disruption rate of star
clusters and the intergalactic interaction history. In this sense the star
clusters in the LMC may give hints to how the intergalactic interaction affects
the evolution of star cluster systems \citep{Renaud13}.

Despite being the nearest interacting system of galaxies, the Magellanic
System still has only a small angular fraction observed to the photometric depth
of its old main sequence turn-off (MSTO). Deep observations suggest that the LMC
stellar populations extend beyond an angular distance of 15$^{\circ}$ from its
centre \citep{Majewski09}. There is also kinematic evidence for a
dynamically warm stellar component consistent with a halo \citep{Minniti03} that
has its major axis oriented with the disk \citep{Alves04}.

Very few studies are available in the outskirts of the LMC.
\citet{WeinbergNikolaev01} report an exponential scale length of $R_s \sim 1.4$
kpc with no significant distinction between a young and old disk; however, their
sample is from the 2 Micron All-Sky Survey \citep[2MASS;][]{2MASS}, which is
very shallow and does not allow for a clear age selection.
\citet{Saha10} report a smaller scale length of $R_s = 1.15$ kpc based on
an optical survey. They argue in favour of a truncation radius of $R_t \sim 14$
kpc. Their sample is limited to only a few fields and their analysis does not
use age selected stellar samples.

A new generation of photometric surveys \comm{is} now coming to the Southern
Hemisphere, allowing for the first time a complete view of the Magellanic
System. One of them is the Dark Energy Survey (DES), which will observe the
outskirts of the LMC, SMC, and most of the Magellanic Stream over the course of
5 years. DES is a photometric survey with the primary goal of
measuring the dark energy equation of state. To achieve this goal the survey
will employ four independent cosmological probes: galaxy clusters, baryon
acoustic oscillations, weak lensing, and type Ia supernovae \citep{Flaugher05}.
The total survey area is $\sim$ 5000 deg$^2$ reaching a magnitude limit
of $i \sim 24$. The photometric system adopted for DES comprises the filters
$g,r,i,z$, which are similar to the ones used in the Sloan Digital Sky Survey
\citep[SDSS;][]{Fukugita96}, with the addition of the $Y$ passband, which provides synergy with the
VISTA Hemisphere Survey \citep{VHS, Cioni11}.  The DES footprint will overlap
with several other surveys in the southern hemisphere allowing a multi
wavelength approach to various astrophysical problems.

An early release of DES Science Verification (SV) data was made available to the
DES collaboration recently. The data cover $\sim$200 deg$^2$ of the southern sky
and sample regions as close as 4$^{\circ}$ North from the LMC centre. The
photometric catalogue from this release reaches $\sim$ 1.5 mag fainter than the
old MSTO of the LMC (which is at $g\sim22$), allowing for a detailed study of
the resolved stellar population of this galaxy.

In this paper we use the DES SV data to study the LMC geometry and
density profile as traced by stellar components with a characteristic age range.
We model the distribution of stars using a simple projected exponential disk and
perform a formal fit. We discuss the presence of a possible truncation radius
and its implication for the LMC mass. As an alternative probe of the LMC
geometry we use Red Clump (RC) stars as a distance indicator and a ruler for the
LMC thickness. This paper is organized as follows. In Section 2 we present a
brief introduction to the DES SV data. In Section 3 we discuss the quality of the
photometry and address issues due to completeness and survey coverage. In
Section 4 we describe the disk model and the fitting procedure used to find the
geometrical parameters of the LMC. Section 5 shows our efforts to use the RC as
a distance and thickness estimator. In Section 6 we summarize and discuss the
implications of the results found in this paper.

\section{DECam and DES science verification}

The Dark Energy Camera (DECam) \citep{Flaugher10} was constructed in
order to carry out the Dark Energy Survey. This instrument has a focal plane
comprised of 74 CCDs: 62 2k$\times$4k CCDs dedicated to science imaging and 12
2k$\times$2k CCDs for guiding, focus, and alignment. The camera is also equipped
with a five element optical corrector and a sophisticated cryogenic cooling
system. DECam is installed at the prime focus of the Cerro Tololo Inter-American
Observatory (CTIO) 4 meter Blanco telescope. In this configuration, DECam has a
2.2 degree-wide field-of-view (FoV) and a central pixel scale of 0.263
$\arcsec$/px.  In typical site conditions, the Blanco telescope plus DECam yield
an imaging point spread function (PSF) with full-width half maximum (FWHM) of
$0.9\arcsec$, which is adequately sampled by the pixel scale.

DECam was commissioned in September 2012 and began operations in November 2012,
with the DES SV campaign covering a period of three months. The DES SV data is
intended to test capabilities of the camera, the data transfer infrastructure,
and the data processing. All images taken during SV are public;
however, the catalogues generated by the collaboration are proprietary.

\subsection{Data reduction}

The DES data management (DESDM) team was responsible for the reduction of the
SV images. Here we will give a brief description of the reduction process. For
a complete description we refer to \citet{Sevilla11}, \citet{Desai12}, and
\citet{Mohr12}.

The DESDM data reduction pipeline consists of the following steps:
\begin{description}
    \item[\bf Image detrending:] this step includes correction for crosstalk
        between CCD amplifier electronics, bias level correction,
        correction for pixel-to-pixel sensitivity variation
        (flat-fielding), as well as corrections for non-linearity,
        fringing, pupil, and illumination.
    \item[\bf Astrometric calibration:] in this step, bright known stars are
        identified in a given image using \textsc{SExtractor} \citep{Bertin.sex}. The
        position of these stars in the focal plane is used to find the
        astrometric solution with the aid of the software \textsc{SCAMP}
        \citep{Bertin.scamp} through comparison to UCAC-4 \citep{UCAC}.
    \item[\bf Nightly photometric calibration:] several reference stars are
        observed each night, and a photometric equation is
        derived for that night. This equation takes into account a zero point, a
        colour term, and an airmass term for each of the DECam science CCDs.
    \item[\bf Global calibration:] relative photometric calibration of the DES
        survey area is done with repeated observations of the same star in
        overlapping DECam exposures using a method similar to that described in
        \citet{Glazebrook94}. The relative magnitudes are anchored to a small
        set of absolutely calibrated reference stars within the same area called
        tertiary standards which come from observations on photometric nights.
        They are then calibrated relative to known equatorial belt standards
        observed on the same night with the same filter set \citep{Tucker07}.
        The current accuracy of the relative and absolute systems is a few
        percent, and is expected to improve as the DES survey covers larger
        contiguous areas. \comm{DES calibration was checked using the Star Locus
        Regression technique \citep{Kelly14}. This analysis revealed that
        zero-point offsets are typically less than 0.05 mag in $g$, $r$, $i$, and
        $z$ across the DES footprint observed so far.}
    \item[\bf Coaddition:] in order to increase the signal-to-noise ratio,
        exposures are combined. This has the advantage of mitigating
        transient objects, such as cosmic rays and satellite trails. This step
        requires the placement of the images in a common reference projection.
        The software \textsc{SWarp} \citep{Bertin.swarp} was used and the
        coaddition process was done in segments of the sky called \emph{tiles}.
        At this stage the flux is corrected according to the photometric
        calibration described above.
    \item[\bf Cataloging:] for each coadded \emph{tile} source detection
        and model-fitting photometry is performed using \textsc{PSFEx}
        and \textsc{SExtractor} \citep{Bertinpsf, Bertin.sex} on a combined $r,
        i,$ and $z$ image. Object fluxes and many other characteristics are
        calculated in the individual $g,r,i,z,Y$ frames. These catalogues
        are ingested into a high performance database system.
\end{description}

The final catalogue is available for the DES collaboration through a database
query client. There are approximately 900 parameters measured for each source
identified by SExtractor. In Table \ref{tab:pars} we list a few parameters
relevant to this work. In the same table we also list any comments about the
parameter and the quality cuts applied.

\begin{table}
\caption{Here we show a subset of the \textsc{SExtractor} parameters measured for the
coadded DECam SV data. We also show any selection criteria that were made.}
\label{tab:pars}
\begin{tabular}{lc}
\hline
Parameter name & Selection \\
\hline
\tt{RA}                       &    --                       \\
\tt{DEC}                      &    --                       \\
\tt{MAG\ul AUTO\ul *}         & $\le$ 26.0 in $g$ and $r$   \\
\tt{MAGERR\ul AUTO\ul *}      &         --                  \\
\tt{SPREAD\ul MODEL\ul *}     & $|$\texttt{SPREAD\ul MODEL\ul I}$| \le 0.002$ \\
\tt{FLAGS\ul *}               & $\le$ 3 in $g$ and $r$      \\
\hline
\end{tabular} \\
\medskip 
The $*$ symbol refers to all passbands available.
\end{table}

\section{The SV data}

By the end of the SV campaign, DECam had obtained images in five passbands of a
region with roughly 300 deg$^2$, for which 200 deg$^2$ are contiguous. The
contiguous region covers the northern outskirts of the LMC. This
region overlaps an eastern portion of the South Pole Telescope (SPT)
footprint \citep{Carlstrom09}. Hence we call it SPT-E for simplicity.

In this section we will discuss several aspects of the SV data and how it is
suitable for the analysis we propose.

\subsection{Photometry}
\label{sec:phot}

Currently, the deepest and most homogeneous magnitude measurement resulting from
the DESDM pipeline is \texttt{MAG\ul AUTO}, which is computed using the flux
inside a Kron radius \citep{Kron80}. The Kron radius is dependent on how
extended a source is, hence the aperture is variable, but it is essentially the
same for all stars, \comm{given that there are no significant spatial dependence
in the image quality, which is the case for DES observations.} Thus,
\texttt{MAG\ul AUTO} is roughly equivalent to a simple aperture magnitude but
less sensitive to seeing variations. We adopt the following notation throughout
the paper: $g$ is the {\tt MAG\ul AUTO} magnitude measured for the $g$ passband.
This also applies for the magnitudes measured in $r, i, z, Y$.


The raw DESDM catalogue has approximately $10^{8}$ sources with $g \le 24.6$.
This list includes spurious detections like satellite trails, star wings, cosmic
rays, etc. To exclude such detections from our catalogue we adopt a simple cut
in the {\tt FLAGS} parameter. We select only sources that simultaneously have
{\tt FLAGS\ul G} and {\tt FLAGS\ul R} $\le 3$, which selects objects that
are not saturated and do not contain any bad pixel. The \texttt{FLAGS} code is
the same as the one adopted by \textsc{SExtractor}. The number of sources left
after this cut is $\sim 9\times10^7$.

We check the stability of the photometric calibration provided by DESDM by
comparing the RC peak colour at different points of the SPT-E that contain LMC
stars. For all bands we found a maximum scatter of 0.02 mags around the RC peak
colour.

The following model was adopted to describe the photometric uncertainties from
the DECam SV data. 
\begin{equation}
    \label{eq:photerrors}
    \sigma(mag) = a + \exp \left( \dfrac{mag - b}{c} \right)
\end{equation}

\begin{figure}
    \includegraphics[width=0.45\textwidth]{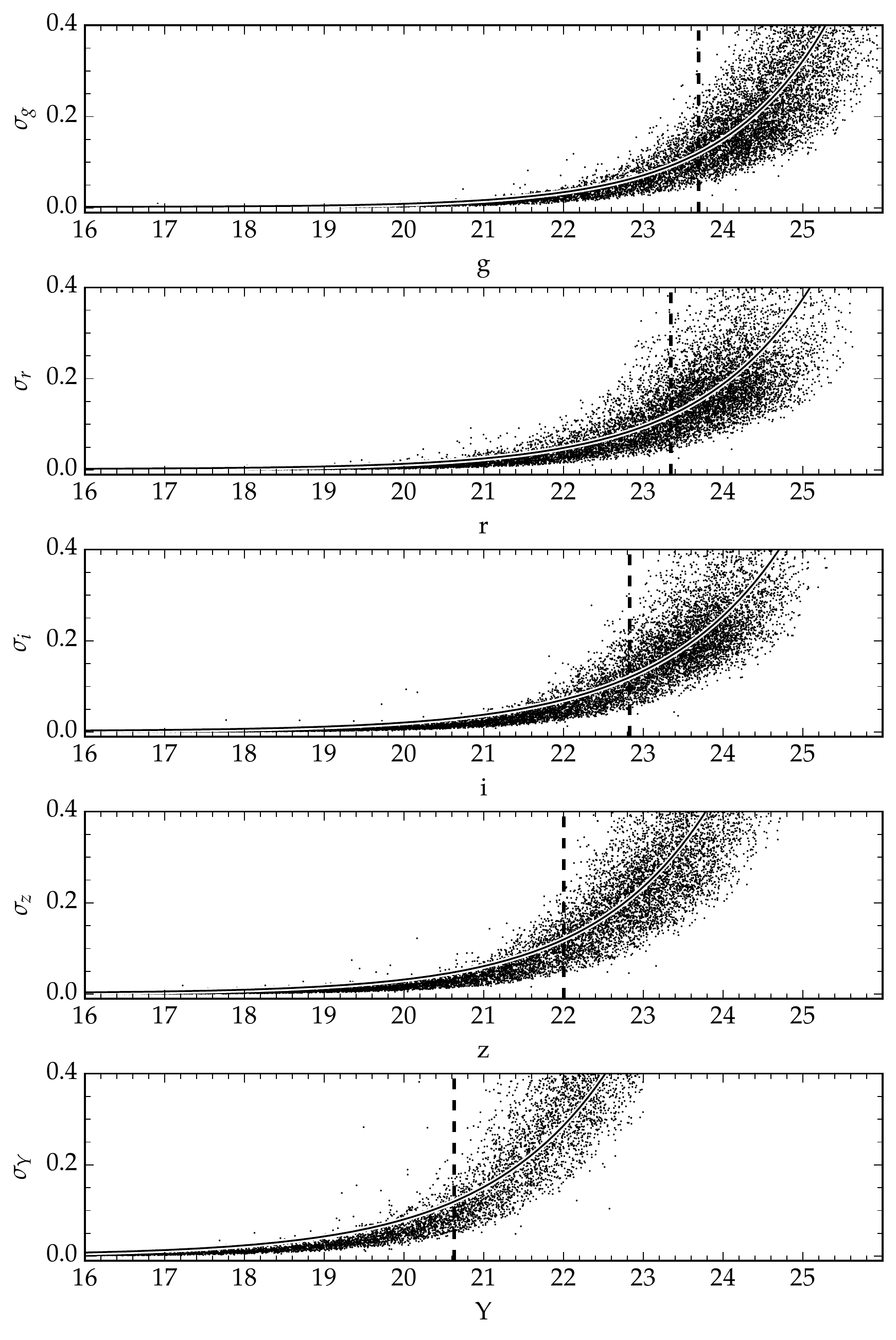}
    \caption{Magnitude versus uncertainty for a sample of 0.01\% of the
    total number of stars in the DECam SV region. The solid curve shows the
    best-fit error model. The coefficients are given in Table \ref{tab:photerrors}.
    The \comm{dashed} line shows the 10\% uncertainty magnitude.}
    \label{fig:photerrors}
\end{figure}

By fitting the above equation to a sample of 0.01\% randomly chosen stars from
the SPT-E region we find the error curves shown in Figure \ref{fig:photerrors}.
The coefficients for each curve are given on Table \ref{tab:photerrors}, where
we also show the 10\% uncertainty magnitude for each band.  Along with
these values we also give some basic information about the filter central
wavelength and extinction coefficients at those wavelengths from
\citet{Cardelli89}.

\begin{table}
\caption{Some general photometric system information: the central
    wavelength ($\lambda_c$) in nanometres, the extinction as a fraction of the
    extinction in the Johnson V passband (assuming $R_V = 3.1$ and a
    \citet{Cardelli89} extinction curve for the MW), and the value of the
    magnitude corresponding to a typical signal-to-noise ratio of 10. In the
    last three columns the coefficients of the best-fit error model are shown.}
\label{tab:photerrors}
\begin{tabular}{lcccc}
\hline
Filter & $\lambda_c$ & $A_{\lambda}/A_V$ & $m_{10\%}$ & $(a$, $b$, $c)$ \\
       & (nm) & & & \\
\hline
$g$    &  479    & 1.199  &  23.94 & (0.001, 26.41, 1.25) \\
$r$    &  641    & 0.837  &  23.76 & (0.001, 26.34, 1.27) \\
$i$    &  781    & 0.635  &  22.75 & (0.003, 25.52, 1.34)  \\
$z$    &  924    & 0.462  &  22.03 & (0.003, 24.75, 1.43)  \\
$Y$    &  1008   & 0.400  &  20.50 & (0.009, 23.45, 1.40)  \\
\hline
\end{tabular}
\end{table}

\subsection{Star/Galaxy Separation}

The DES collaboration has generated several datasets for validating and testing
the DESDM system \citep{Mohr12}. These datasets were simulations of the actual
observations. To create these \emph{mock} observations, input catalogues of
artificial galaxies and stars were generated \citep{Rossetto11, Balbinot12,
Busha}. The mock observations included a time varying seeing, realistic
shapes for the galaxies, and variations to the focal plane of DECam. These
images were fed into the DESDM reduction pipeline and an output catalogue was
generated, which was then released to the collaboration to perform tests of
their scientific algorithms.

Tests with a set of such simulations, called Data Challenge 5 (DC5) by the
collaboration, have been carried out in order to assess how different
star-galaxy classification parameters perform. A summary of this comparison is
shown in \citet{Rossetto11}. The main conclusion is that \texttt{SPREAD\ul
MODEL} \citep{Desai12, Bouy13} performs better in terms of purity and
completeness than other typically employed classifiers such as \texttt{CLASS\ul
STAR} and \texttt{FLUX\ul RADIUS} \citep{Bertin.sex}. Later,
\citet{Soumagnac13} developed a more sophisticated star-galaxy separation
algorithm; however, this algorithm must be trained on a data set where
true stars and galaxies are known.  Implementation of such methods are being
considered collaboration wide. For simplicity, we choose to use
\texttt{SPREAD\ul MODEL} measured in the $i$ band as the star-galaxy separator.
The cut-off criterion for selecting stars is $|$\texttt{SPREAD\ul MODEL\ul I}$|
\le 0.002$. From test on DC5, this cut is found to correspond to a simultaneous
stellar completeness and purity of $\simeq 80\%$ for objects with $g < 23$
\citep{Rossetto11}. From this point further we call objects that match
this cut-off criteria \emph{stars}. It is worth mentioning that at $g=23$ we
expect $\sim20\%$ of contamination from galaxies in our sample. However, the
large scale distribution of these objects is homogeneous and is unlikely
to significantly affect the findings of this paper.

\subsection{Completeness}
\label{sec:comp}

To independently assess the completeness of the DESDM catalog we conducted a few
experiments using \textsc{DAOPHOT} \citep{Stetson87}. This photometry code is
known to perform very well in extremely crowded regions such as the cores of
globular clusters \citep{Balbinot09}. Hence, at the typical density of LMC field
stars it should yield a fairly complete catalogue. This \textsc{DAOPHOT}
catalogue may be compared to the one produced by DESDM as an approximation to a
complete catalogue and giving an estimate of the completeness as 
function of magnitude. This is not the most accurate approach to this problem;
however, it is much simpler than performing artificial star experiments across
several hundred squared degrees.

\begin{figure}
    \includegraphics[width=0.45\textwidth]{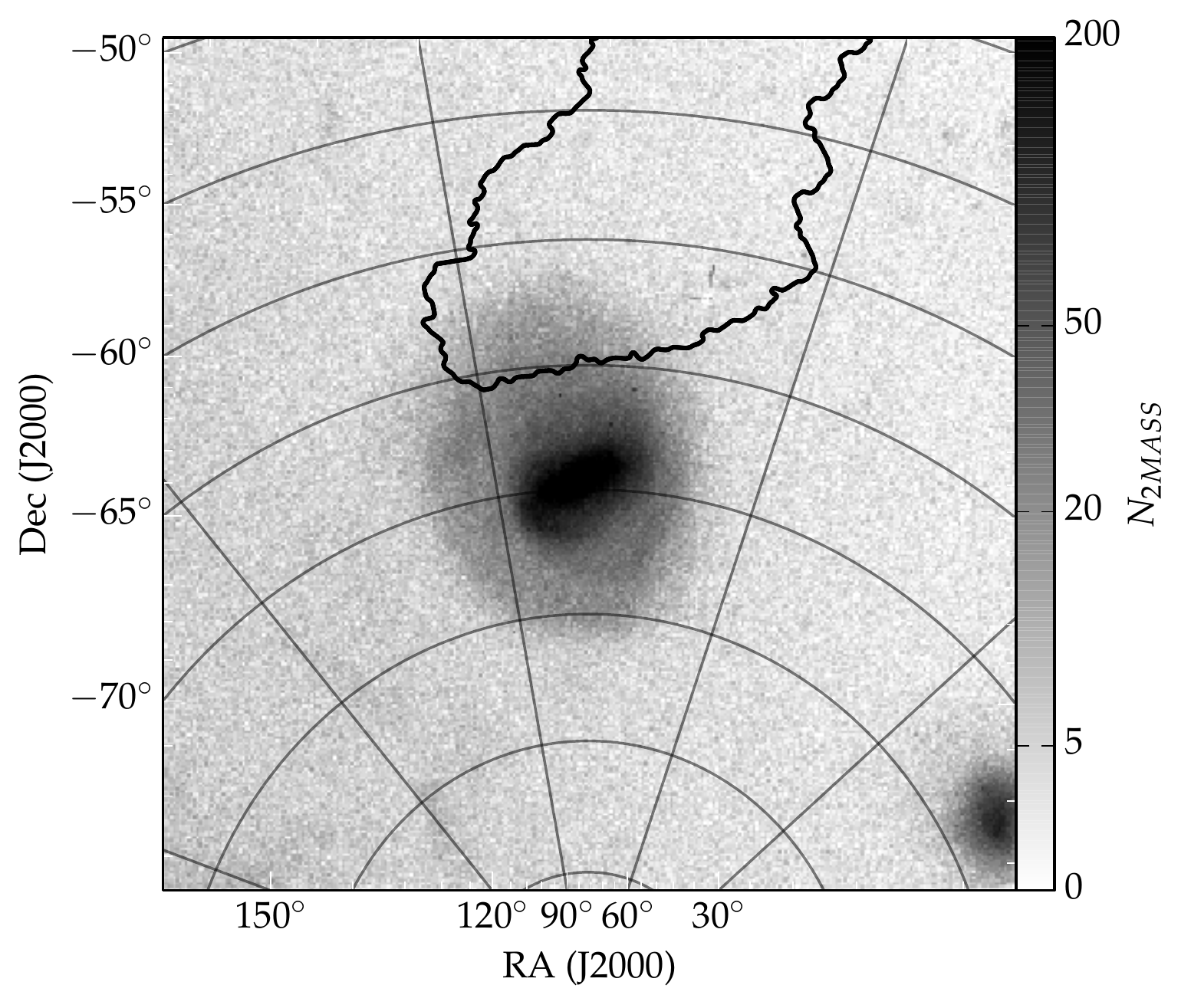}
    \includegraphics[width=0.45\textwidth]{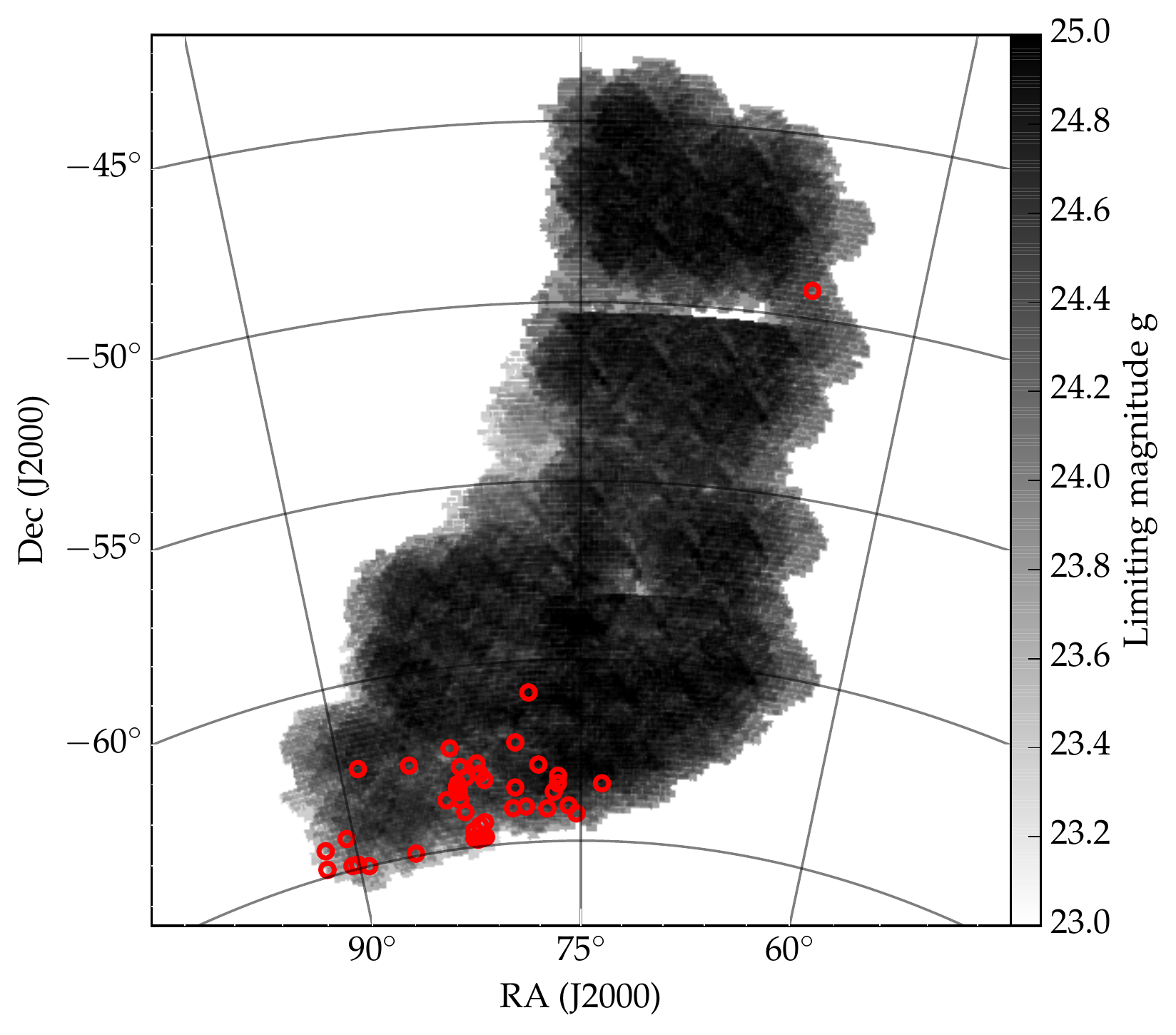}
    \caption{Top panel: Gnomonic projection of the number of 2MASS stars that
        are consistent with AGB and RGB stars according to \citet{Yang07}. The
        projection is centred in the LMC centre. Part of the SMC is visible in
        the lower right corner. The solid contour shows an approximate footprint
        of the southern part of the SPT-E region of the DES SV data. Bottom
        panel: Gnomonic projection of a $N_{side}$=4096 \textsc{HEALpix} map of
        the $g$ magnitude limit \textsc{Mangle} mask for the SPT-E region quoted
    above. Holes and unobserved regions are masked and shown in white. Red
circles mark the position of the fields selected for assessing the survey
completeness.}
    \label{fig:mangleg}
\end{figure}

\comm{
We performed \textsc{DAOPHOT} photometry on 51 fields, 50 of which contained a
LMC star cluster in its centre. The 51$^{\rm st}$ field  was selected as far
away as available in the SPT-E data in order to sample a region where there are
few or no LMC stars. The fields selected are subregions of the coadded images
encompassing $6.75\arcmin\times6.75\arcmin$ each, with the exception of the
51$^{\rm st}$ field, which has $18\arcmin\times36\arcmin$. The first 50 fields
were selected in order to assess the completeness not only in regions with a
density of stars typical of the LMC, but also with varying density, such as the
inner regions of a star cluster. A broader discussion about this subject will be
presented in a future paper. The 51$^{\rm st}$ field was selected in order to
compare the performance of the DESDM reduction pipeline in a region with little
or no crowding. The position of each of the all 51 fields is marked on Figure
\ref{fig:mangleg}. On average DESDM and DAOPHOT photometry agree within 0.02 in
$g$ and $r$.}


To compare the number of stars as a function of magnitude, we cut both the
DESDM and \textsc{DAOPHOT} catalogues at the 3\% photometric uncertainty
level. This cut happens at $g\simeq23.5$. To separate stars from galaxies in
\textsc{DAOPHOT} we adopted a cut in the \emph{sharpness} parameter which
behaves similarly to \texttt{SPREAD\ul MODEL}.

\begin{figure}
    \includegraphics[width=0.45\textwidth]{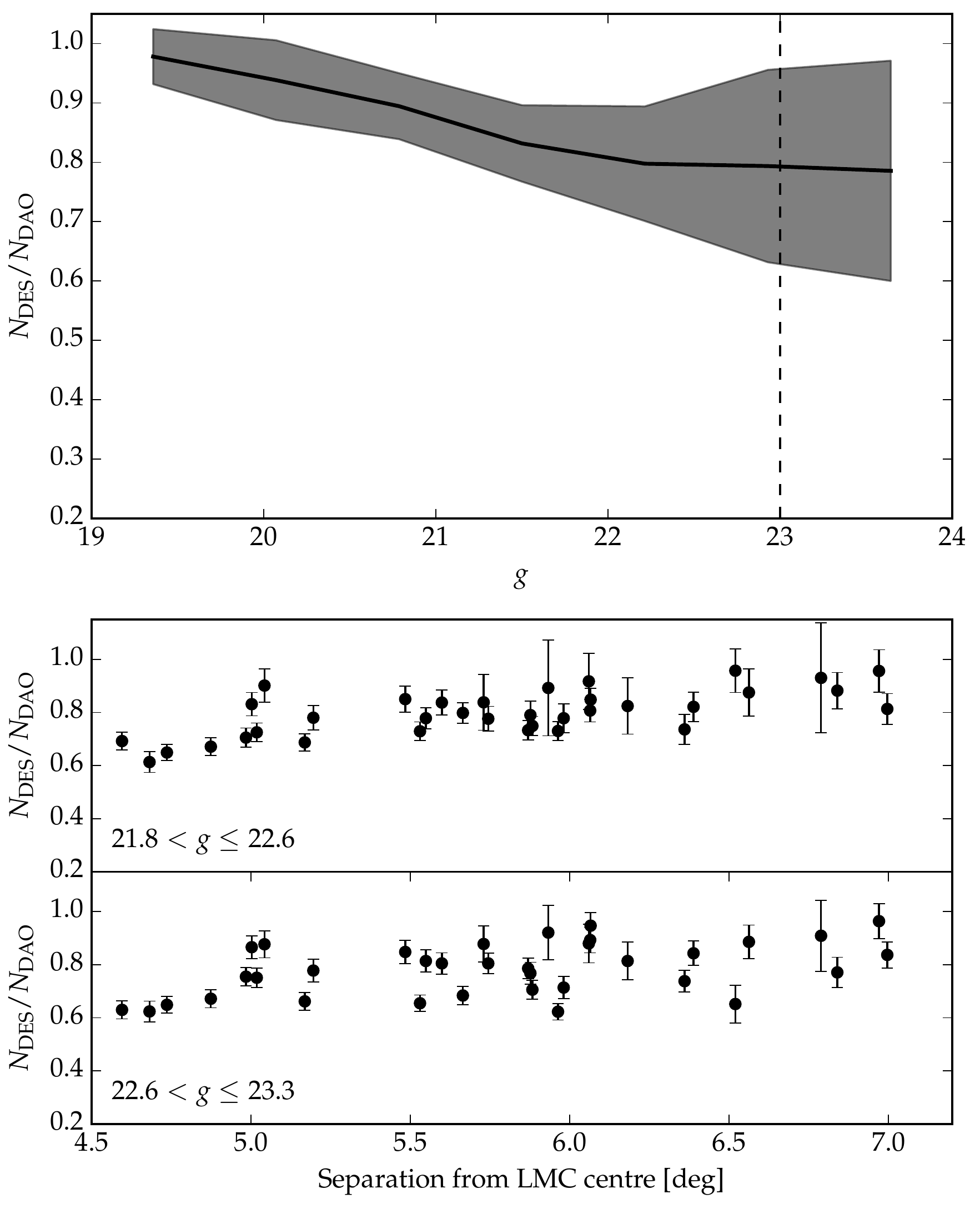}
    \caption{\comm{\emph{Top panel:} we show the ratio of stars detected by DESDM and
        \textsc{DAOPHOT} as a function of $g$ magnitude. The solid line shows
        the average for the 51 fields analysed. The shaded region shows the
        standard deviation for the 51 fields. The dashed vertical line shows the 
        faintest magnitude limit used in this work. \emph{Bottom panels:} black
        circles show the same ratio as in the top panel but now as a function of
        angular separation to the LMC centre for two magnitude bins (indicated
        in the captions). The error bars are computed using the Poisson
        uncertainty.}}
    \label{fig:completeness}
\end{figure}

\begin{figure}
    \includegraphics[width=0.45\textwidth]{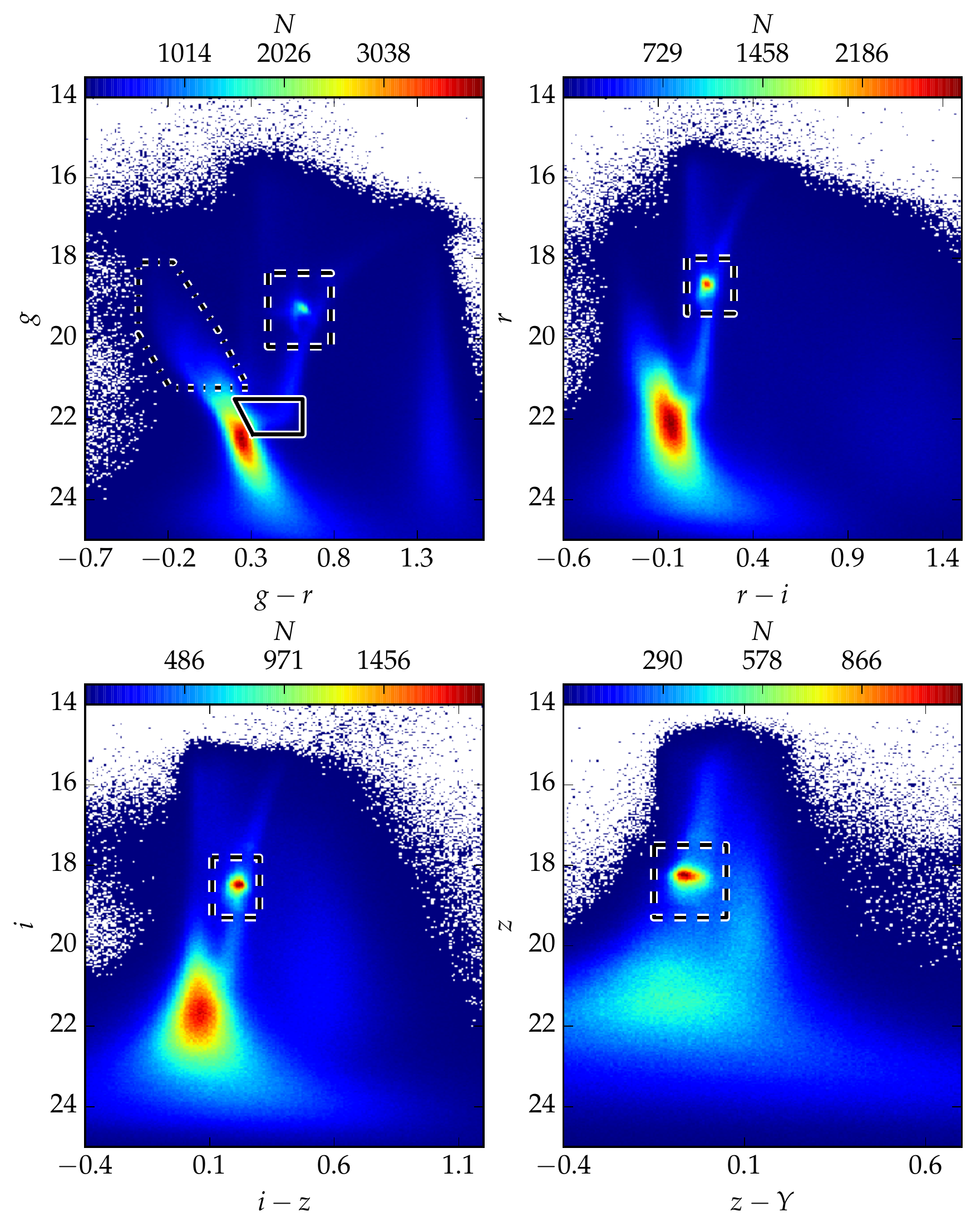}
    \caption{The Hess diagrams in each set of colour and magnitude available in
        the DES photometric system. In the top left panel we also show the
        colour-magnitude polygons used to select stars consistent with a
        young/intermediate (dot dashed polygon) and old (solid polygon) stellar
        population. The boxes (dashed) used for the selection of RC stars are shown
        in all panels.}
    \label{fig:hess}
\end{figure}

\comm{
In Figure \ref{fig:completeness} top panel we show the ratio between the number
of stars detected by DESDM and \textsc{DAOPHOT} ($N_{DES}/N_{DAOPHOT}$) averaged
in magnitude bins for the 51 fields discussed above. The solid line shows the
average value at a given magnitude bin. The shaded region shows the standard
deviation. In the two bottom panels in the same figure we show the value of
$N_{DES}/N_{DAOPHOT}$ as a function of angular separation to the LMC centre. We
show this for two magnitude bins:  $21.8 < g \le 22.6$ and $22.6 < g \le 23.3$
(also indicated in the figure). From Figure \ref{fig:completeness} we notice
that the average completeness drops as a function of increasing magnitude, as
expected. The completeness remains roughly constant as a function of angular
separation to the LMC centre with a spread of the same order as the one shown in
the shaded region of the top panel. We conclude that the completeness has little
spatial dependence at least up to magnitudes as faint as $g\sim23$.
}


Despite the good performance in moderately crowded fields (i.e. LMC field
population), we observed that in the inner regions of star clusters the DESDM
sample is highly incomplete reaching less than 40\% completeness for $g=19$ and
dropping to zero for fainter magnitudes. \comm{This will be further explored in
future publications}.

We note that where the completeness is most important to our results regarding
truncation radii and scale lengths, i.e. at distances farther away from
the center of the LMC, the star density is low enough that crowding effects
do not affect the sample selection as shown in Figure
\ref{fig:completeness}.

From the completeness analysis described above we conclude that in the DESDM
catalogue completeness does not depend strongly on source density, except
in extremely crowded environments such as the central parts of stars clusters.
From visual inspection \comm{and through comparison with state-of-the-art crowded
field photometry methods} we find no strong evidence for large-scale
variations in the stellar completeness in the SPT-E region, thus allowing us to
properly access the spatial distribution of stars across its footprint, either
from the LMC or from the MW.

\subsection{Mangle masks}

We use the \textsc{Mangle} software \citep{Swanson08} to track the survey
coverage and limiting magnitude. Along with the release of the DESDM
catalogues, a set of \textsc{Mangle} masks was also provided. These masks contain
information about the limiting magnitude in each patch of the sky, as well as
other key information about data quality. The magnitude limit is computed as the
detection limit of a point source with signal-to-noise ratio $S/N=10$. The noise
is estimated from the variance and level of the sky on a patch of the sky. To
measure the flux an aperture of 1$\arcsec$ radius is used; this yields a
magnitude measurement called \texttt{MAG\ul APER\ul 4}. For a complete
description of the masks we refer to \citet{Swanson12}.

In Figure \ref{fig:mangleg} we show the number of AGB and RGB stars found in
2MASS according to \citet{Yang07} with the southern limits of the SPT-E region
overplotted (top panel). We also show an approximation of the magnitude
limit \textsc{Mangle} mask for the SPT-E region (bottom panel). It is an
approximation in the sense that it uses the value of the \textsc{Mangle} mask at
the central position of each pixel in the sky. The gray scale represents the
magnitude limit at each point of the footprint. Holes caused by bright stars or
other image imperfections are displayed as well. These regions have a  magnitude
limit of zero, hence going out of the gray scale range. The white color in the
figure represents regions that were not observed. The figure uses a
\textsc{HEALpix} pixelization \citep{hpx} with $N_{side}$=4096 and a Gnomonic
projection centered at $(\alpha, \delta) = (75^{\circ}, -55^{\circ})$.

The approximate mask discussed above allows us to deal with the full SPT-E mask
in a much faster way and at the same time to provide a very good approximation
of the general properties of the survey, such as coverage and magnitude limits.
The coverage mask is simply a \textsc{HEALpix} map with the value of 1 for
pixels that were observed and 0 for those that were not or were masked for some
reason. The coverage mask originally did not contain holes due to star clusters.
To mask these regions we conducted a visual search for star clusters in the
SPT-E and used the position of each cluster to add a hole in the mask (i.e. a
region with coverage equals to zero).

The approximated masks were used to remove stars residing in regions
where the following criteria were met: (i) a limiting $g$ or $r$ magnitude
brighter than 23; (ii) coverage value equal to 0. The additional trimming for
zero mask values was necessary because the original catalogue was limited by
the \emph{exact} masks, not the approximated ones.

The above section outlines the process of trimming the catalogue to select
objects that are likely stars and to keep only regions with deep photometry that
were not affected by artifacts in the survey. This catalogue, as well as the
mask associated with it, will be used in the remainder of this paper.

\section{The LMC geometry}
\label{sec:geo}

Classically, the LMC is classified as an Irregular Dwarf Galaxy, although it has
several major components of a spiral galaxy such as a disk and a bar. It is,
perhaps, more appropriate to classify this galaxy as a highly perturbed spiral
galaxy. Its morphology departs so much from a classical Irregular Dwarf that it
has been established that the LMC is the prototype of a class of dwarf galaxies
called Magellanic Irregular \citep{deVaucouleurs72}. These galaxies are
characterized by being gas-rich, one-armed spirals with off-centre bars.

The SMC is the closest neighbour to the LMC. Together they form the Magellanic
System. Recent dynamical modelling \citep{Besla12} and high precision tangential
velocity measurements \citep{Kallivayalil13} point to the need for updated
thinking with regard to the origins of the Magellanic System. The centre of
mass spatial velocity of these galaxies is very close to the escape velocity of
the MW, thus suggesting that the system is not gravitationally bound to the MW.
The same models also predict the formation of the Magellanic Bridge and Stream
as a result of the interaction of the LMC with the SMC, generating long arms of
debris in the same fashion as the Antennae system.

Despite the growing number of simulations and high precision velocity
measurements, a few aspects of the LMC geometry, such as the presence of a
spheroidal halo \citep{Majewski09}, and the warping and flaring of the disk
\citep{Subramaniam09} remain uncertain to some degree. New large area surveys in
the Southern Hemisphere are beginning to shed light on these uncertainties.

Here we study the LMC disk geometry using a very simple approach. We try to
model its stellar density using a circular exponential disk. This disk is
inclined relative to the sky plane by the angle $i$. To compute the
expected number of stars $\rho$ as a function of $\alpha$ and $\delta$ we use
the transformations found in \citet{WeinbergNikolaev01}. For the sake of
clarity, we give the expression for the heliocentric distance $t$ to a given
point of the disk with coordinates $(\alpha, \delta)$.

\begin{align}
    \label{eq:weinberg}
    t(\alpha, \delta) &= -R_{LMC} \cos{i} \times \{\cos{\delta}\sin{(\alpha - \alpha_0)}\sin{\theta}\sin{i} \nonumber \\
      &+ [\sin{\delta}\cos{\delta_0} - \cos{\delta}\sin{\delta_0}\cos{(\alpha-\alpha_0)}]\cos{\theta}\sin{i} \nonumber \\
      &-[\cos{\delta}\cos{\delta_0}\cos{(\alpha-\alpha_0)} +
      \sin{\delta}\sin{\delta_0}]\cos{i}\}^{-1}
\end{align}
where $(\alpha_0, \delta_0)$ is the central coordinate of the LMC, $i$ is the
disk inclination, $R_{LMC}$ is the heliocentric distance to the LMC centre, and
$\theta$ the position angle (PA) of the minor axis. In the reference frame adopted
here, the inclination $i$ has negative values when the North side of the LMC is
closer to us. In the reference frame adopted, the $i$ angle is
reversed to what is typically adopted in the literature.

The density of stars is simply given by

\begin{equation}
    \rho(\alpha, \delta) = \rho_{0} t(\alpha, \delta)^2exp(-R/R_{s}) + \rho_{BG} f(\alpha,\delta)
\end{equation}
where $R$ is the radial distance in the disk plane, $R_{s}$ is the scale length
of the exponential disk, $\rho_0 t(\alpha_0, \delta_0)$ is the central density,
and $\rho_{BG}$ is the density of background/foreground stars. The
function $f(\alpha, \delta)$ is a third degree polynomial which takes into
account the spatial variation of MW field stars.

Five parameters are used to model the disk geometry: the LMC central coordinates
$(\alpha_0, \delta_0)$, its distance to the Sun ($R_{LMC}$), the inclination
$i$, and the PA $\theta$. Additionally, 3 parameters describe the density of
stars along this disk: the central density $\rho_{0}$, the scale radius $R_s$,
and the density scale of background stars $\rho_{BG}$.

To simplify the problem and to better accommodate the fact that the
observations are all on the northern side of the MCs, we make a few
assumptions. The LMC centre is kept fixed at the \citet{Nikolaev04} value of
$(\alpha_0, \delta_0) = (79.40^{\circ}, -69.03^{\circ})$. The heliocentric
distance to the LMC centre $R_{LMC} = 49.9$ kpc is adopted from the most recent
review about the subject \citep{deGrijs14}.

\begin{figure}
    \includegraphics[width=0.45\textwidth]{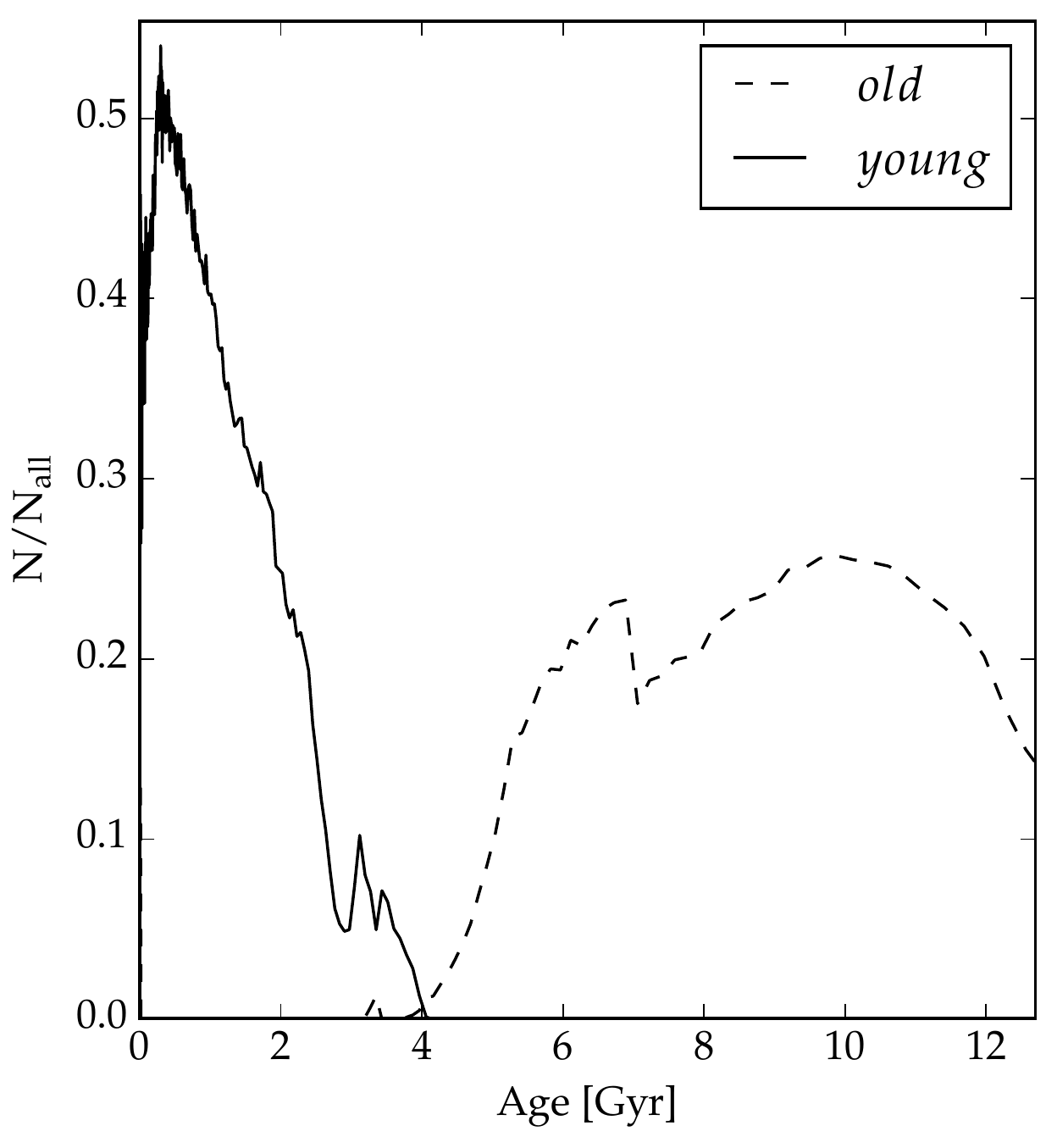} 
    \caption{Ratio of the number of \emph{young} (\emph{old}) to \emph{all}
        stars as a function of age. The ratio was computed using a constant SFR
    and a \citet{Piatti12} AMR.}
        \label{fig:popratio}
\end{figure}

With the machinery to produce LMC disk models we proceed to a formal fit to the
observed distribution of stars in the SPT-E region. This fit was performed on
three samples of stars. The selections were made based on \textsc{PARSEC}
\citep{Bressan12} stellar evolution models. The first is made only of stars
older than $4$ Gyr, selected using a colour-magnitude cut shown in Figure
\ref{fig:hess}. The second sample selects only stars younger than $4$ Gyr. The
third sample is made of stars that fall within the limits $17 < g < 23$ and
$-0.5 < g-r < 1.0$. We name these samples \emph{old}, \emph{young}, and
\emph{all} for simplicity. \comm{In Figure \ref{fig:popratio} we show the ratio
    of \emph{old} (\emph{young}) and \emph{all} stars for a simulated LMC
    stellar population that assumes a constant star formation rate (SFR) and a
    \citet{Piatti12} age-metallicity relation (AMR). Photometric errors were
    simulated according to Equation \ref{eq:photerrors} and Table
    \ref{tab:photerrors}. We observe that the colour-magnitude cuts chosen are
able to create two samples with virtually no age overlap.}

Using a \textsc{TRILEGAL} \citep{Girardi05} simulation, we modeled the expected
MW stellar population in the SPT-E region. Assuming the same colour cuts as the
ones used to select the three stellar populations quoted above, we fit the
density of stars as a function of RA and Dec using a third order polynomial,
$f(\alpha, \delta)$. The best-fit polynomial for each of the stellar populations
cuts is later used to account for foreground contamination while fitting the LMC
disk. This polynomial is scaled by $\rho_{\rm BG}$ since the colour-magnitude
cuts are not perfectly consistent with the theoretical prediction of
\textsc{TRILEGAL}.

\begin{figure*}
    \begin{minipage}{\textwidth}
    \includegraphics[width=0.45\textwidth]{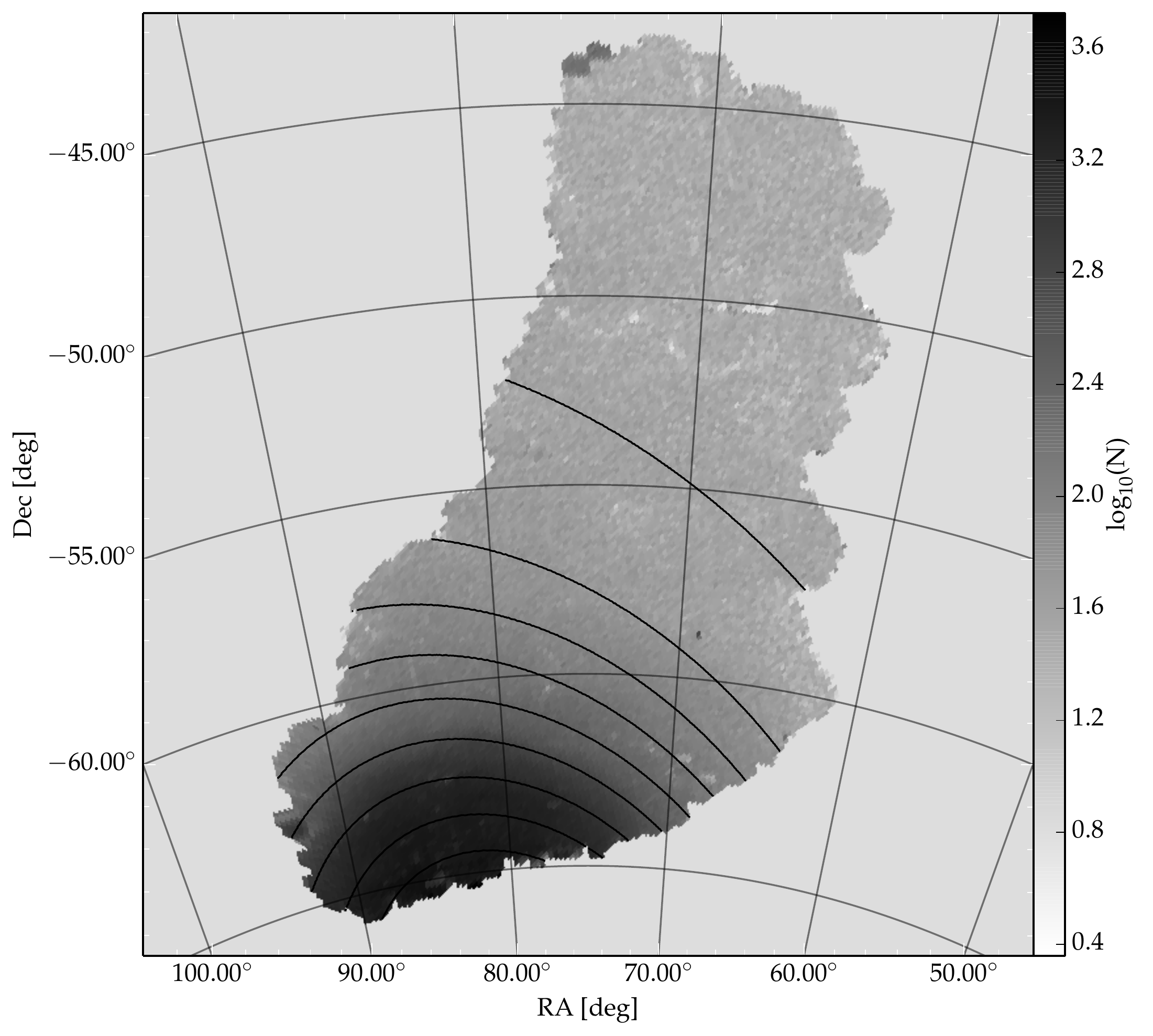}
    \includegraphics[width=0.5\textwidth]{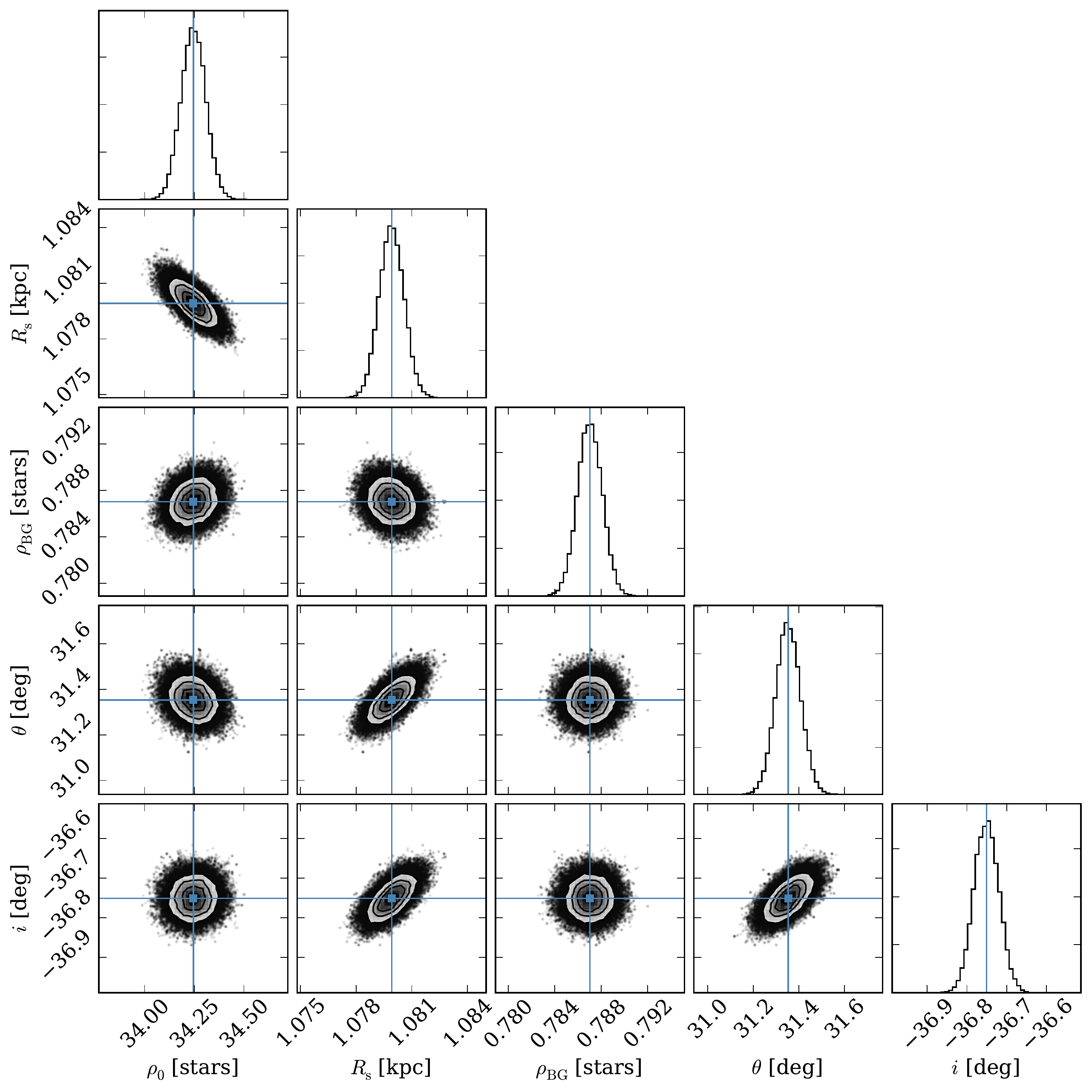}
    \caption{\emph{Left panel:} map showing the number of stars from the
        \emph{all} sample in grayscale. The stars used to build this map are
        those that fall inside a simple color cut of $-0.5 < g-r <1.0$ and $g <
        23$. The solid isodensity contours show the best exponential disk model.
        The contours start at $\log_{10}(N) = 3.3$ and progress in steps of 0.3
        dex. \emph{Right panels:} the marginalization for the different disk
        model pairs of parameters. The solid blue crosshair shows the
        position of the best solution. The contours show the 1, 2, and 3$\sigma$
        confidence levels. The histograms in the diagonal panels show the
        marginalization over each single parameter. Again the blue line shows the point
        of maximum likelihood.}
    \label{fig:disk_all}
    \end{minipage}
\end{figure*}

\begin{figure*}
    \begin{minipage}{\textwidth}
    \includegraphics[width=0.45\textwidth]{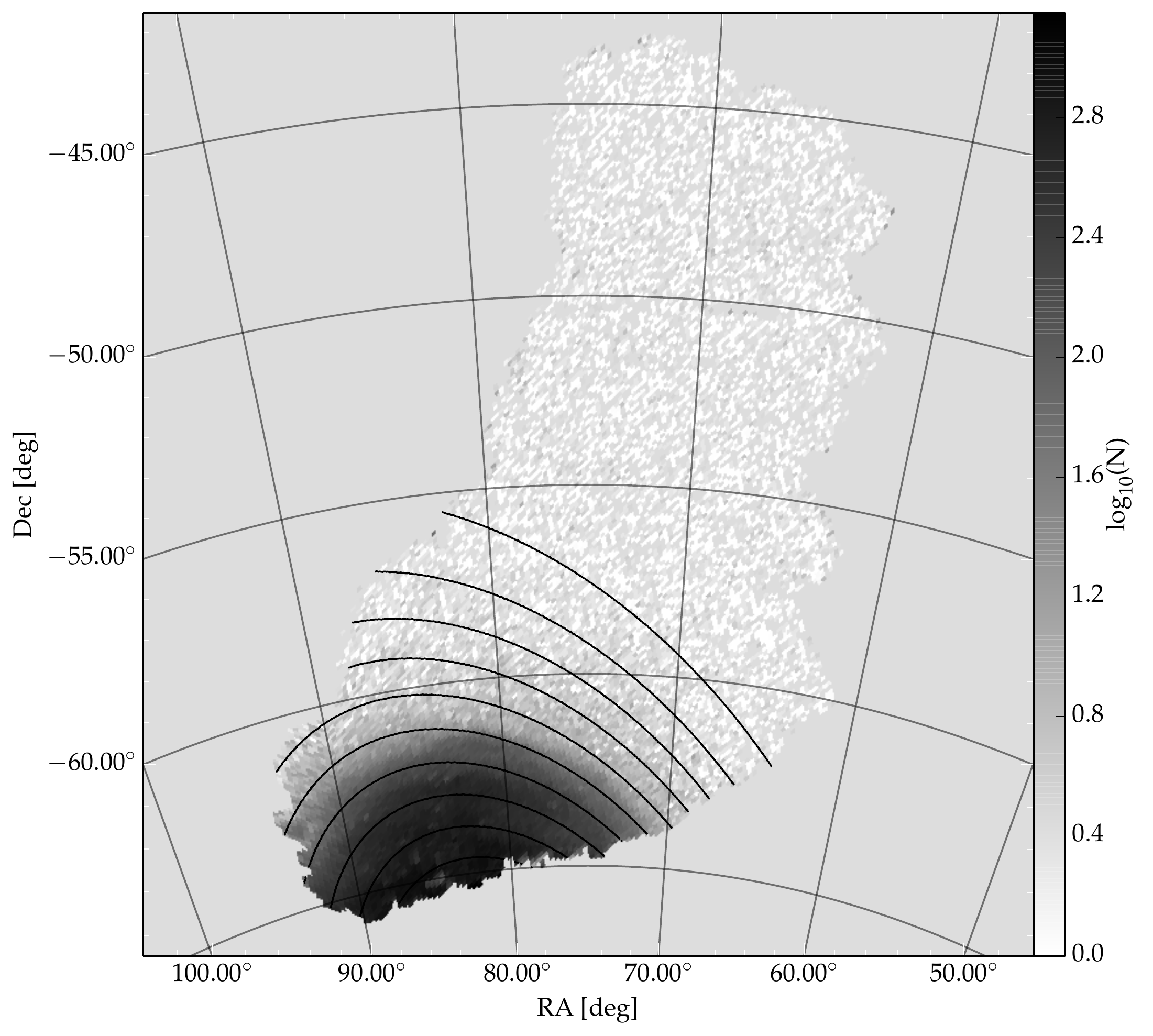}
    \includegraphics[width=0.5\textwidth]{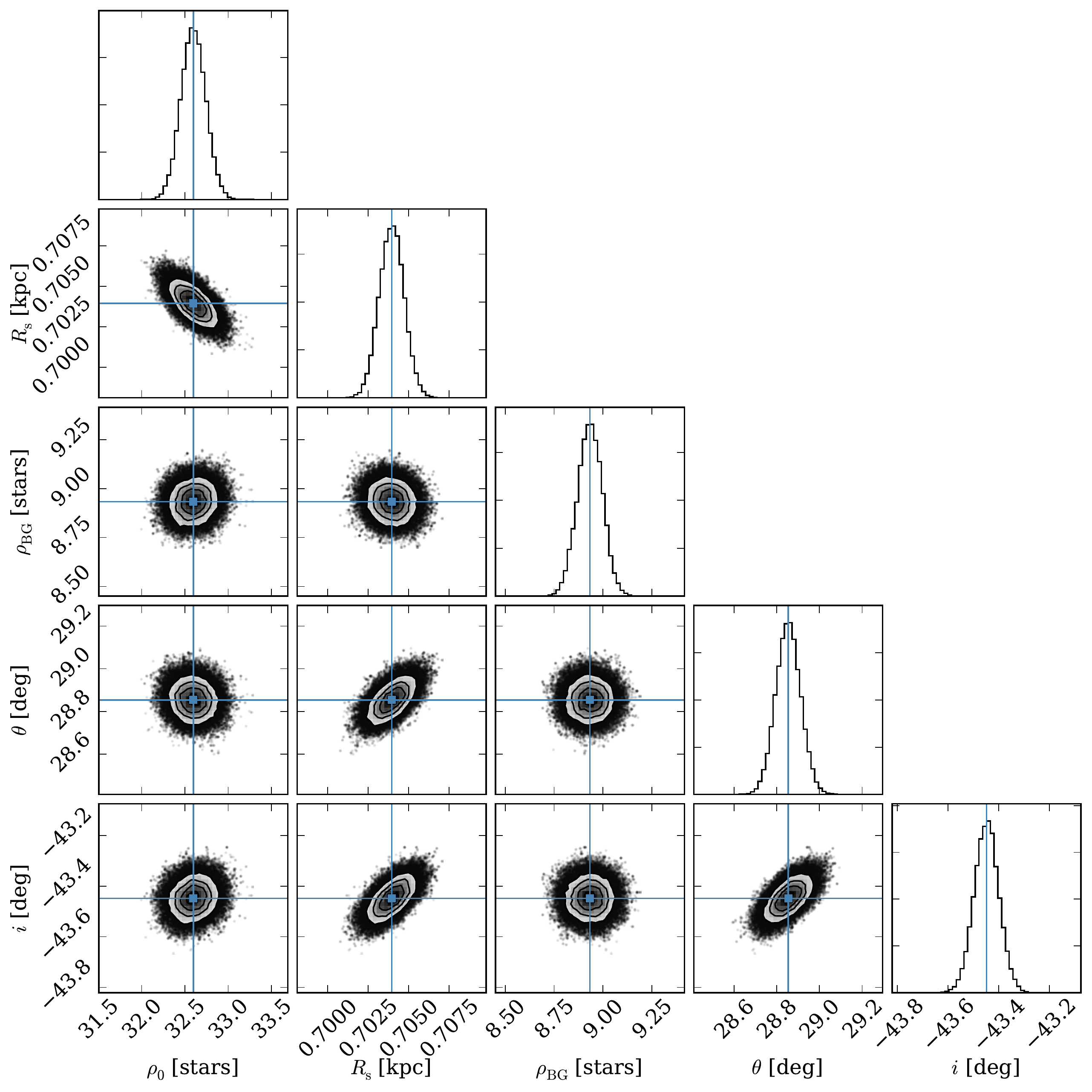}
    \caption{Similar to Figure \ref{fig:disk_all} but for the \emph{young}
    stellar population. The isodensity contours start at $\log_{10}(N) = 2.9$
and progress in steps of 0.4 dex.}
    \label{fig:disk_young}
    \end{minipage}
\end{figure*}

\begin{figure*}
    \begin{minipage}{\textwidth}
    \includegraphics[width=0.45\textwidth]{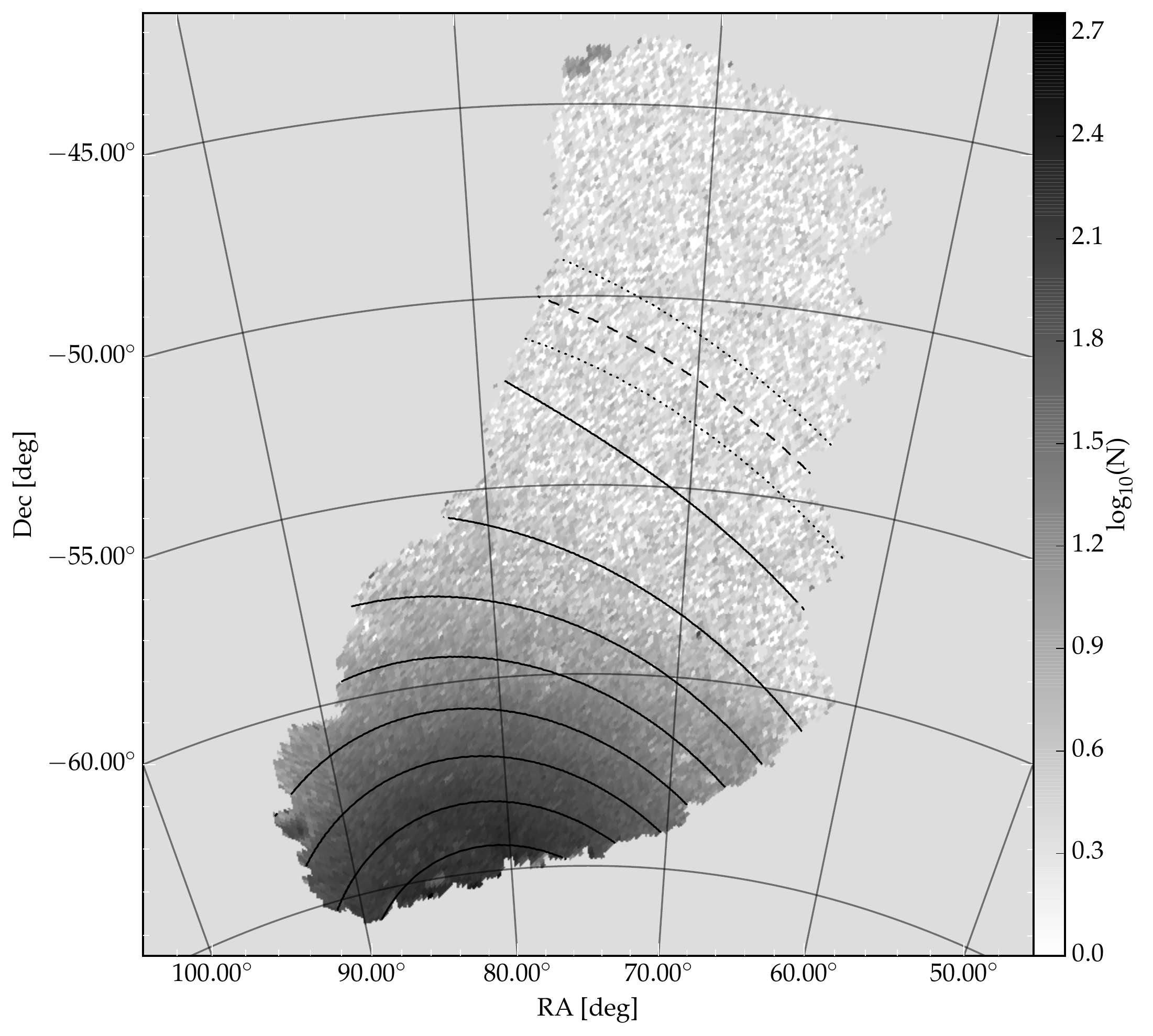}
    \includegraphics[width=0.5\textwidth]{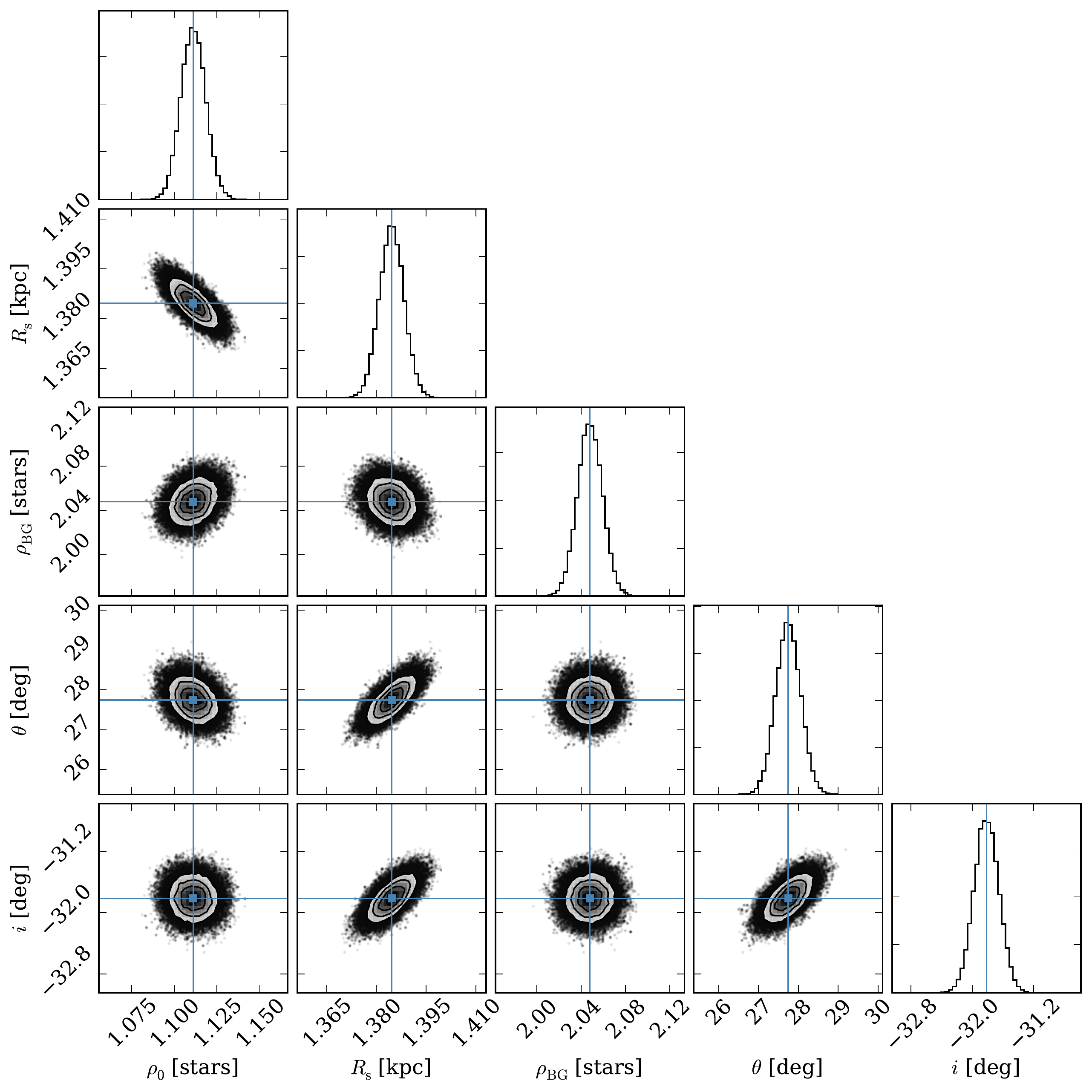}
    \caption{Similar to Figure \ref{fig:disk_all} but for the \emph{old}
    stellar population. The isodensity contours start at $\log_{10}(N) = 2.5$
and progress in steps of 0.35 dex. In the left panel we also show the truncation
radius ($R_{\rm t}$) position as the dashed contour, and the dotted contours show the
uncertainty in $R_{\rm t}$.}
    \label{fig:disk_old}
    \end{minipage}
\end{figure*}

To count stars in the sky plane we adopt the \textsc{HEALpix} pixelization
scheme with $N_{side} = 512$ which yields a constant pixel area across the
sky of $\sim 0.013$ deg$^2$. The maps built in this scheme for each stellar
sample described above are shown in the left panel of Figures
\ref{fig:disk_all}, \ref{fig:disk_young}, and \ref{fig:disk_old}.

The best-fit disk model was obtained through a Markov Chain Monte Carlo (MCMC)
technique. We used the code \textsc{emcee} \citep{Foreman-Mackey13} in its
version 2.0.0. The test statistics chosen is a binned Poisson log-likelihood
model \citep{Dolphin02}. We refer the reader to these authors for details
on the MCMC and statistics used. This is the most appropriate test statistics
for data-model comparison since we are dealing with counts that are subject to
shot noise, especially in the outermost regions of the LMC. Our MCMC run uses a
total of 30 walkers that make 1000 steps each for the \emph{burn-in} phase.
After the \emph{burn-in}, we let the walkers advance 5000 more steps each,
sufficient to well sample the parameter space and converge to the maxima.

This fitting procedure was repeated for each stellar population (\emph{all, old
} and \emph{young}), yielding the parameters shown on Table \ref{tab:best_fit}.
In this table we chose to present the PA of the line
of nodes $\theta_0 = \theta+90^{\circ}$. This is the quantity most often
presented in the literature.

\begin{table*}
\begin{minipage}{\textwidth}
\centering
\begin{tabular}{lcccccc}
\hline
Population & Age & $\theta_0$ & $i$ & $R_s$ & $\rho_0$      & $\rho_{BG}$ \\
           & Gyr & deg        & deg & kpc   & stars pixel$^{-1}$ kpc$^{-2}$ & \\
\hline
All  & --    & $129.51\pm0.17 (\pm1.08)$ & $-38.14\pm0.08 (\pm1.59)$ & $1.09\pm0.01(\pm0.02)$ & $26.41\pm0.14 (\pm 0.59)$ & $0.80\pm0.03  (\pm 0.67)$\\
Young& 0 - 4 & $125.93\pm0.20 (\pm0.09)$ & $-44.19\pm0.14 (\pm1.80)$ & $0.72\pm0.01(\pm0.00)$ & $22.86\pm0.28 (\pm 1.33)$ & $9.06\pm0.18  (\pm 0.00)$\\
Old  & 4 - 13& $127.40\pm1.02 (\pm0.59)$ & $-32.94\pm0.39 (\pm1.25)$ & $1.41\pm0.01(\pm0.00)$ & $0.89 \pm0.02 (\pm 0.05)$ & $2.08\pm0.04  (\pm0.01)$\\
\hline
\end{tabular}
\caption{A summary of the best-fit disk models parameters for each stellar
    population. The uncertainties are the 3$\sigma$ confidence level that arises
    from the MCMC analysis. The uncertainties in parenthesis are obtained
    from the difference between the fit with and without correcting the
    magnitudes by the heliocentric distance to each point of the LMC disk.
    NOTE: our definition of inclination ($i$) has the opposite sign than what is
    typically found in literature.}
\label{tab:best_fit}
\end{minipage}
\end{table*}

The boxes chosen to select a given stellar population in the colour-magnitude
diagram (CMD) assume that the stars are all at the same distance to us. This is
not strictly the case for the LMC stars, which are spread across a disk plane
inclined relative to the sky plane. To test how this distance spread affects
our disk fit we correct the magnitude of each star using the best-fit disk
models so as to bring the stars to a common distance. This common distance is
chosen as the mean distance to the disk in the SPT-E. Using these distance
corrected magnitudes we perform the CMD box selections again.  For all the CMD
based selections an increase of $<1\%$ in the number of stars inside the CMD
box is observed. Using this distance corrected sample we rerun the fitting
procedure and find that the largest change in the parameters is a 5\%
increase in the inclination. The remaining disk parameters are more stable.
We adopt the difference on the parameters found in this experiment as the
systematic uncertainty in our parameter estimation. We show this
variations as the uncertainties in parenthesis in Table
\ref{tab:best_fit}.

The left panel of Figure \ref{fig:disk_all} a \textsc{HEALpix} map in
Gnomonic projection shows the logarithm of the number of stars for the
\emph{all} stellar population. Overploted we show the best-fit disk model as
isodensity lines. On the right side of the figure we show the sampled
probability distribution function for each parameter after marginalization.  In
light blue we mark the point of maximum probability density, which indicates the
maximum likelihood solution. Notice that the samples are distributed in a fairly
symmetric way around the maxima, yielding symmetric error bars. On Figures
\ref{fig:disk_young} and \ref{fig:disk_old} are similar to the figure
described above, but showing different age-selected populations.

We notice that the \emph{old} population spreads out to declinations of
$\sim -55^{\circ}$, while the \emph{young} population is much more abruptly
truncated at $\sim -60^{\circ}$. This points to different scale radii for these
populations. In Table \ref{tab:best_fit} we give a summary of the parameters
that best-fit each case. Here we see that the distributions have significantly
different values of $R_{\rm s}$ while retaining similar values for the
purely geometrical parameters $\theta_0$ and $i$.

\begin{figure}
    \includegraphics[width=0.4\textwidth]{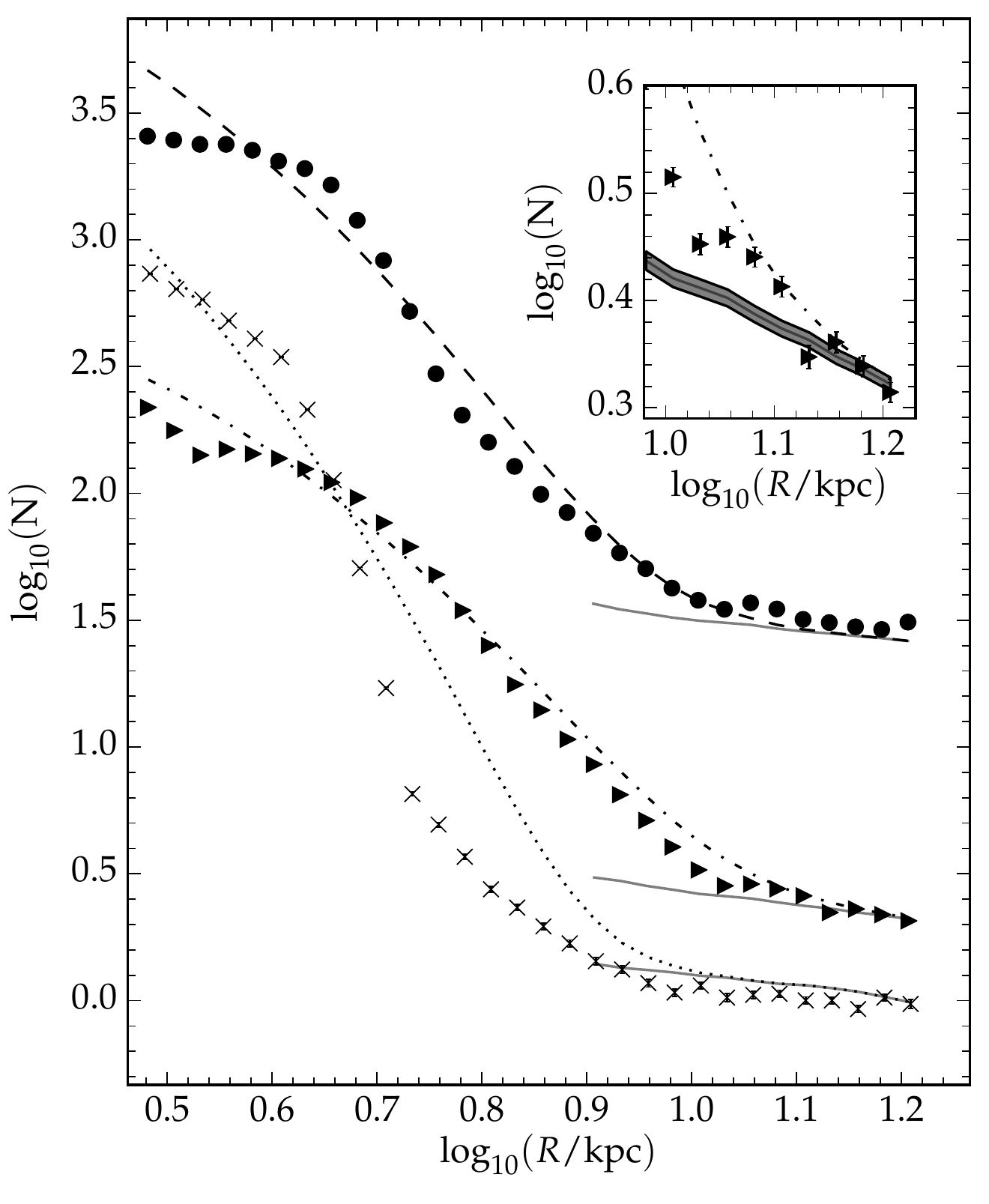}
    \caption{The black circles show the averaged number of stars from the
        \emph{all} sample per \textsc{HEALpix} pixel in different bins of
        distances along the LMC disk. The triangles (crosses) show the same but
        for the \emph{old} (\emph{young}) sample. The dashed, dot-dashed, and
        dotted lines show the best-fit disk model for the \emph{all, old,} and
        \emph{young} populations, including the term $\rho_{BG}f(\alpha,\delta)$
        that accounts for the contamination of MW stars. In the inset plot
            we show the outer tail of the \emph{old} profile. The solid line shows
        the MW contribution and the shaded region shows its uncertainty,
    propagated from the uncertainty in $\rho_{\rm BG}$. The errobar represents
    the Poission uncertainty on the number of counts per bin.}
    \label{fig:profile1d}
\end{figure}

The difference in the scale length is much more obvious in Figure
\ref{fig:profile1d} where we show, for each stellar population, the average
number of stars per \textsc{HEALpix} pixel in bins of distance along the LMC
disk. We also show the best-fit disk model for each of the stellar populations,
this model includes the MW foreground stars (i.e. the term $\rho_{\rm
BG}f(\alpha,\delta)$ in Equation \ref{eq:weinberg}). We notice that the
\emph{old} (triangles and dot-dashed line) profile is much more extended than
the \emph{young} one (crosses and dotted line). It is also remarkable that the
LMC \emph{old} density profile is well fit without the need for other
components such as a spheroidal halo. However, the \emph{young} profile
is not very well described by the disk model. This could also account for the
value found for its inclination, which is not in good agreement with the
literature \citep{vdMarel01, Nikolaev04, Rubele12}.  In the same figure, the
solid line shows the contribution of MW stars, we notice that the slope in the
number of stars at the outskirts of the LMC can be accounted by for the spatial
variation in the number of MW field stars. Due to the large number of stars in
each radial bin, the Poisson uncertainty is very low. These uncertainties are
only visible, as error bars, on the inset plot, where the outer parts of the
\emph{old} population profile is shown.

We notice from Table \ref{tab:best_fit} that the statistical uncertainties in
the disk parameters are quite small. This is caused by the relatively poor
description that our disk model gives of the LMC stellar population, especially
the \emph{young} population. In Figure \ref{fig:profile1d} this becomes apparent
where we observe that the models deviate significantly from the observations,
even when the uncertainties are considered. This can point to the case where
there are other systematic effects that were not taken into account. One cause
of such effects might be the spatial variation of the star-galaxy classifier.
However, the classifier adopted here is very stable especially at bright
magnitudes ($g<22.5$), thus being improbable to be a significant source of
error. Also, spatial changes in the completeness of the survey are unlikely to
cause a large change in the number of stars per \textsc{HEALpix} pixel (see
Section \ref{sec:comp}). Evidence indicates that the deviations from the fitted
disk model are real features of the LMC structure.  These features are more
apparent for the \emph{young} stellar population, which is most likely to hold
signs of disk perturbations such as spiral arms.

We use the \emph{old} population profile to probe the total extent of the LMC.
We define the truncation radius as the radius where the observed density profile
reaches becomes indistinguishable from the MW foreground population. In the
inset plot of Figure \ref{fig:profile1d} we observe that for $\log(R/kpc) >
1.13$ the profile can be explained solely by the MW contribution. The
uncertainty on this truncation radius ($R_{\rm t}$) is taken as the size of the
radial bin, which is 0.8 kpc. This yields a $R_{\rm t} = 13.5 \pm 0.8$ kpc.


If we assume that the LMC \comm{luminous component} is tidally truncated by the
MW potential we can use the simple theoretical tidal radius formula
\citep{Binney08}, given by:

\begin{equation}
    R_{\rm t} = d_{\rm LMC} \left( \dfrac{M_{\rm LMC}}{2M_{\rm MW}(d < d_{\rm LMC})}\right)^{\frac{1}{3}} \\
\end{equation}
we find the following relation for the LMC mass ($M_{\rm LMC}$) and the MW mass
($M_{\rm MW}$) encircled within a radius equal to the Galactocentric distance of
the LMC ($d_{\rm LMC}$):
\begin{equation}
    \label{eq:mass}
    M_{\rm MW}(d < d_{\rm LMC}) = 24.5^{+8.8}_{-6.4} \times M_{\rm LMC}
\end{equation}
where the Sun is assumed to be at a distance of $d_{\odot}= 8.0$ kpc from the
MW centre. This yields a $d_{\rm LMC}$ of $49.4\pm2.1$ kpc. The uncertainty in
this value was considered for the result in Equation \ref{eq:mass}. The
distance to the Sun is taken as a compromise between two recent determinations
from \citet{Eisenhauer05} and \citet{Gillessen09}. The adoption of
$d_{\odot}=8.5$ kpc would increase $d_{\rm LMC}$ by $\sim$ $0.03\%$, leading to
an insignificant increase in $R_{\rm t}$. Thus, we choose to disregard the
uncertainty in $d_{\odot}$.

\begin{figure}
    \includegraphics[width=0.46\textwidth]{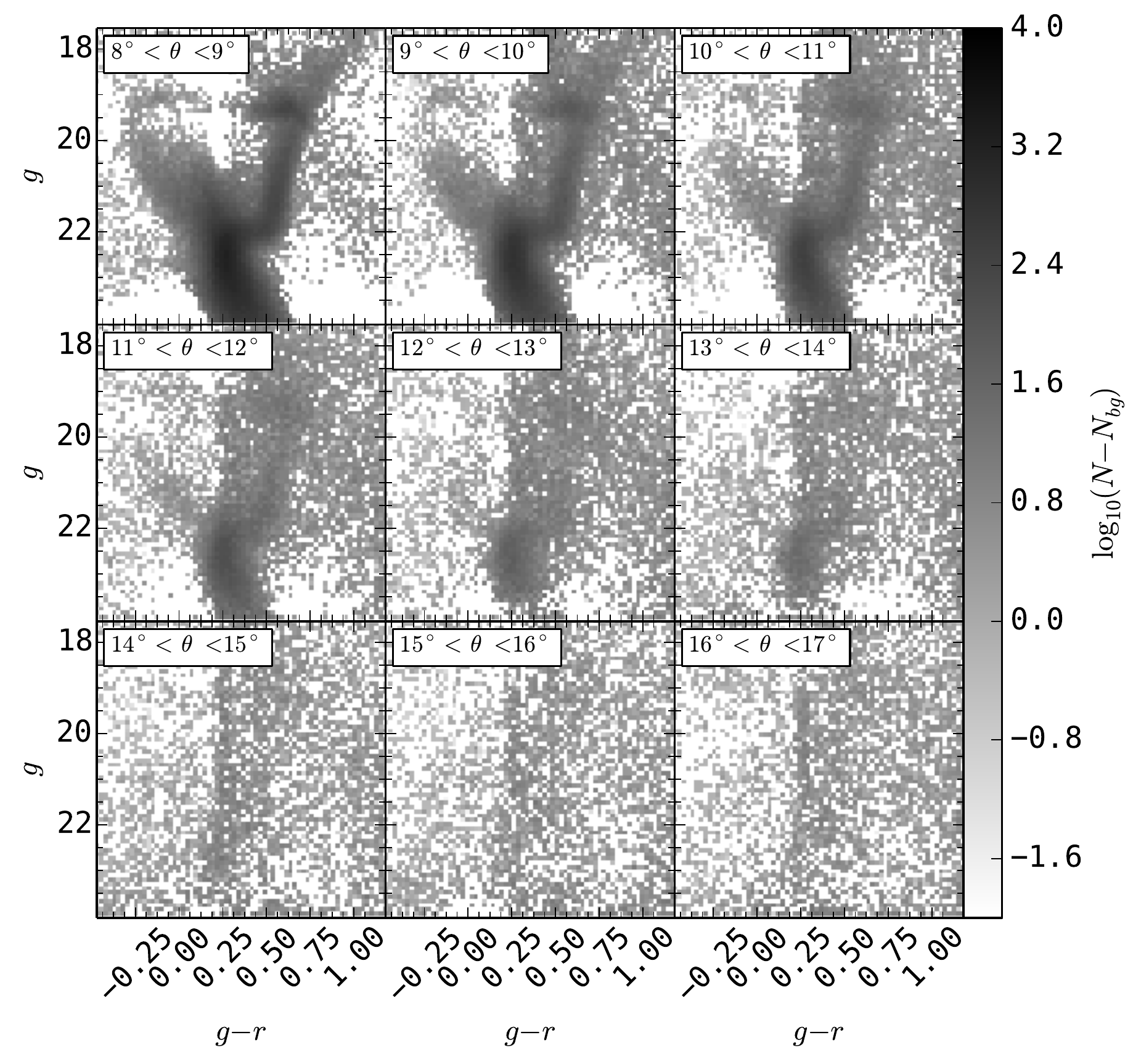}
    \caption{Hess diagrams in $g-r$ vs. $g$ for different angular distance
ranges (indicated in the top left of each panel). The contribution from
MW stars has been removed as described in the text.}
    \label{fig:cmd_spatial}
\end{figure}

To further support our claim that there are very few LMC stars beyond 13 kpc we
show in Figure \ref{fig:cmd_spatial} the Hess diagrams at different bins
of angular separation from the LMC centre. The upper and lower bound of each bin
is indicated in each panel. We apply a simple decontamination algorithm in order
to remove the contribution of MW stars. The decontamination was done by
selecting a region with angular distance greater than $\theta = 20^{\circ}$. A
Hess diagram of this region was constructed. These Hess diagrams were area
weighted and subtracted from the Hess diagram of each angular separation bin
shown in Figure \ref{fig:cmd_spatial}. This process assumes that the
contribution of LMC stars at angular distances larger than $20^{\circ}$ is
negligible. We again assume that the MW stellar population varies very little
throughout the SPT-E.

\comm{In Figure \ref{fig:cmd_spatial} we note, from visual inspection, that the
    LMC population in the bottom central panel is indistinguishable from the
    noise in the Hess diagram. This panel corresponds to angular separations
    between $15^{\circ}$ and $16^{\circ}$. This range corresponds to distances
    from 10.5 and 11.2 kpc along the disk plane, which are intermediate between
    the truncations radius of the \emph{old} and \emph{young} population. This
    does not mean the complete absence of LMC stars. However, it shows that at
    this range of distances the LMC populations is too low in number to be
distinguishable by eye from the MW foreground stars.}

With the aid of the \textsc{galpy}\footnote{http://github.com/jobovy/galpy}
\citep{Bovy10} suite for galactic dynamics we compute a three component MW
potential with a NFW halo \citep{NFW96}, a Miyamoto-Nagai disk \citep{MNdisk},
and a Hernquist bulge \citep{Hernquist90} following the same recipe as in
\citet{Bovy12}. This potential was normalized such as to yield a Solar circular
velocity of 220 km/s at the Galactocentric distance of 8 kpc. We obtain that
the enclosed MW mass inside a sphere centred in the MW with radius $d_{LMC}$ is
$5.6\times10^{11} M_{\odot}$.  Using the result from Equation \ref{eq:mass} and
propagating the uncertainties in $R_t$ we obtain an LMC mass of $M_{LMC} =
2.3^{+0.8}_{-0.6}\times10^{10} M_{\odot}$.

According to the set of simulations by \citet{Kallivayalil13}, the combination
of MW and LMC masses found in this work would favor a scenario where the
Magellanic System is in its first pericentric passage. However, it is not clear
if there is enough time for the LMC to develop a truncation radius without
finishing a complete orbit around the MW \comm{and if this truncation would
affect its luminous component}. We would like to stress that the
calculations presented in our work are subject to many uncertainties, especially
with respect to our knowledge of the MW mass profile at large distances.

\section{The LMC Red Clump as a distance estimator}
\label{sec:rc}

The peak magnitude of the RC is used widely in astronomy as a standard candle
to determine distances of old and intermediate age star clusters. RC
stars are He-burning stars of mass $M \lesssim 2 M_{\odot}$; they develop an
electron degenerate core after the main sequence, and as a consequence have to
increase their core up to a critical mass of $\sim 0.46 M_{\odot}$ before Helium
can be ignited. The almost-constancy of their Helium-core masses determines
that they all share similar (but not constant) luminosities in their central
helium burning phase.

Using the RC peak magnitude as a standard candle we define the distance
modulus $\mu_0$ with the following equation:

\begin{equation}
    \label{eq:dmod}
    \mu_0 = m_{\lambda}^{RC} - M_{\lambda}^{RC} - A_{\lambda} - \Delta M_{\lambda}^{RC}
\end{equation}
where $A_{\lambda}$ is the extinction, $M_{\lambda}^{RC}$ is the absolute
magnitude of the RC peak, and $\Delta M_{\lambda}^{RC}$ is a correction in the
absolute magnitude due to population mixing effects. Here $\lambda$
represents the observational filters ($\lambda=g,r,i,z$).

If one is to determine $\mu_0$, the terms $A_{\lambda}$ and $\Delta
M_{\lambda}^{RC}$ must be inferred. Here we choose to correct the
magnitudes for extinction using the \citet{Schlegel98} dust maps. For the time
being we will assume that $\Delta M_{\lambda}^{RC}$ is zero. This assumption
will be addressed in more detail later.

To probe $\mu_0$ as a function of position in the LMC, we subdivided the sky in
\textsc{HEALpix} pixels using the pixelization scheme with $N_{side}=128$.  This
gives a pixel area of $\sim 0.21$ deg$^2$. We choose a larger pixel than the
previous section since the RC is much less populated than the Main Sequence. To
measure the peak apparent magnitude of the RC in a given passband
($m_{\lambda}^{RC}$) we compute the number count of stars as a function of
magnitude ($N(m_{\lambda})$) for stars that have colours and magnitudes limited
by the dashed boxes in Figure \ref{fig:hess}. The CMD region occupied by
RC stars was selected visually.

To build $N(m_{\lambda})$ the bin size is chosen according to the method described
in \citet{Knuth}, \comm{which is based on optimization of a Bayesian fitness
function across fixed-width bins}. This avoids issues that may arise from
oversampling or undersampling the number of bins. The values of $N(m_{\lambda})$
are fitted by means of a non-linear least square algorithm using a second order
polynomial plus a Gaussian. This function is given by the following equation:

\begin{equation}
    \label{eq:RC_fit}
    N(m_{\lambda}) = a + b m_{\lambda} + c m_{\lambda}^2 + d \exp{\left[\dfrac{-
    (m_{\lambda}^{RC} - m_{\lambda})^2 }{2\sigma_{\lambda}^2}\right]}
\end{equation}
where ${a,b,c}$ are the coefficients of the polynomial, $d$ is the normalization
of the Gaussian and $\sigma_{\lambda}$ is the standard deviation of the
magnitudes around the RC peak.

The uncertainty on $m_{\lambda}^{RC}$ is taken from the covariance
matrix of the least square fit.


It is convenient to define the heliocentric distance to points in the LMC disk
as a function of the so called line of maximum distance gradient. This line
connects points on the disk plane with the most rapidly varying distance from
us, \comm{hence its name}. The distance along this line is the deprojected
distance between a point in the disk to the line of nodes. The distances along
this line are given by the $y$ component of Equation A3 of
\citet{WeinbergNikolaev01}.  \comm{The line of maximum distance gradient} has a
PA given by $\theta$ (Equation \ref{eq:weinberg}) and it is oriented
approximately in the NE-SW direction. \comm{Showing the heliocentric distance to
points in the LMC disk as a function of the position in the line of maximum
distance gradient is the equivalent of showing an edge-on view of the LMC.}

\begin{figure*}
\includegraphics[width=0.45\textwidth]{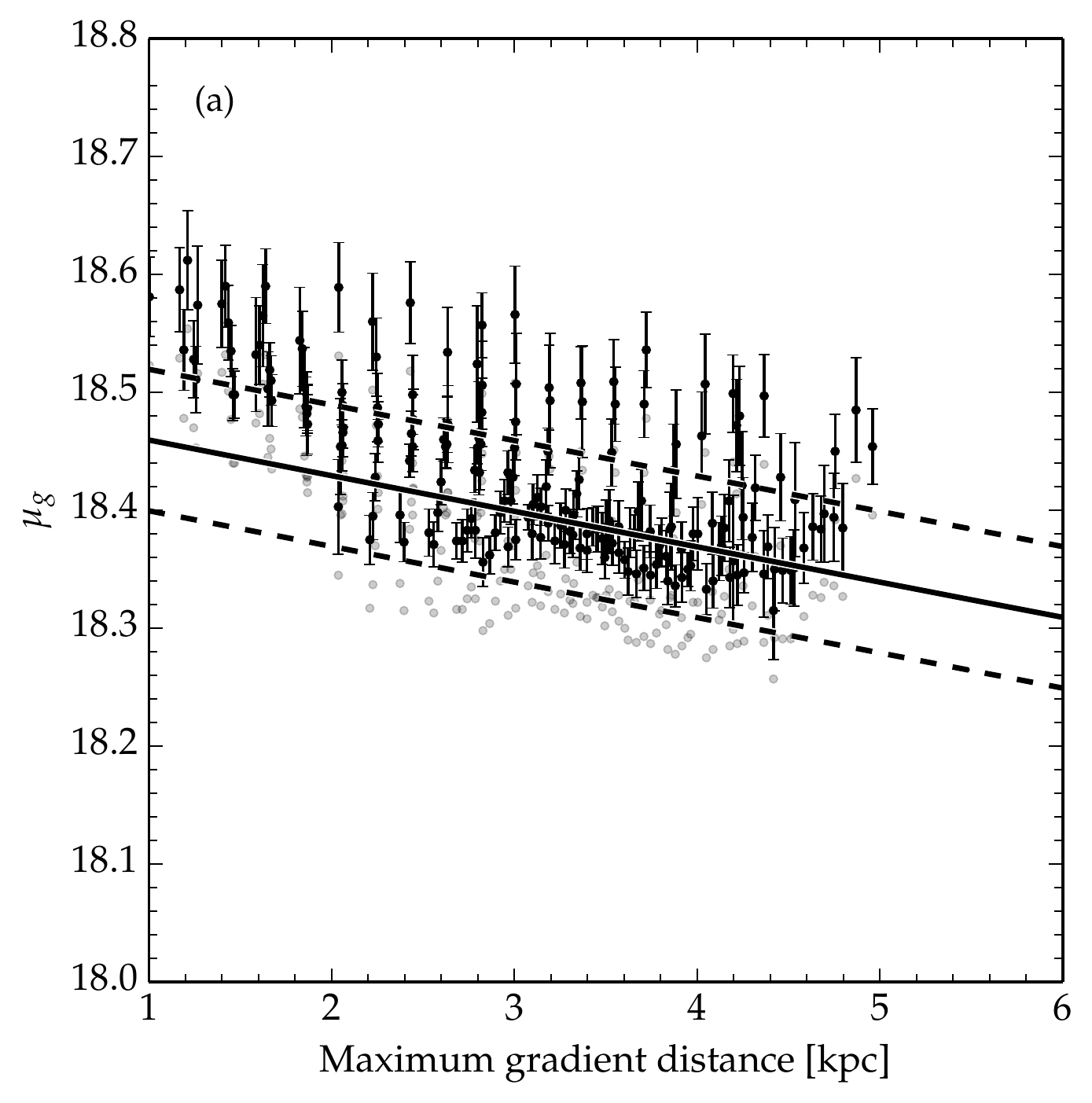}
\includegraphics[width=0.45\textwidth]{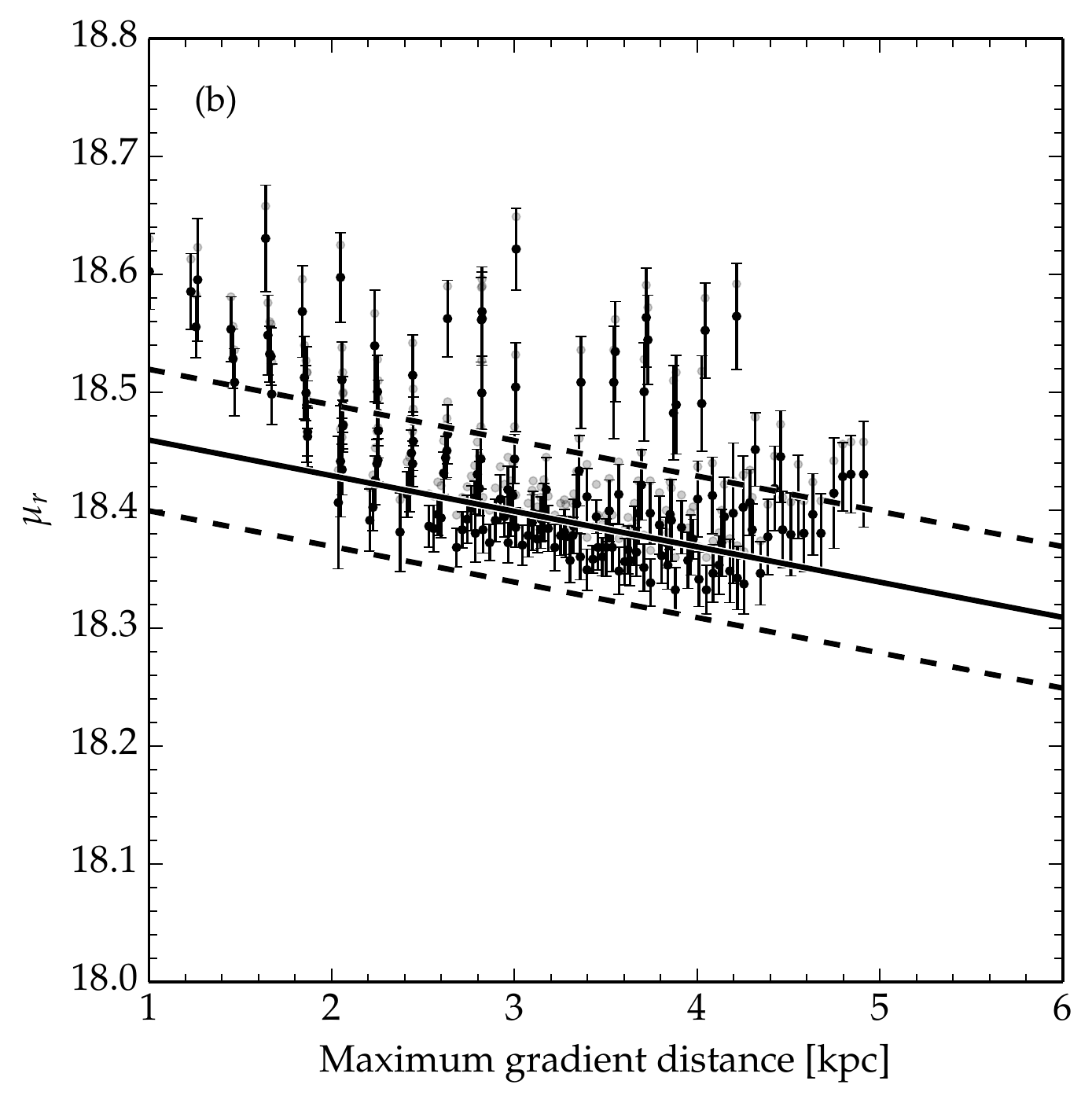} \\
\includegraphics[width=0.45\textwidth]{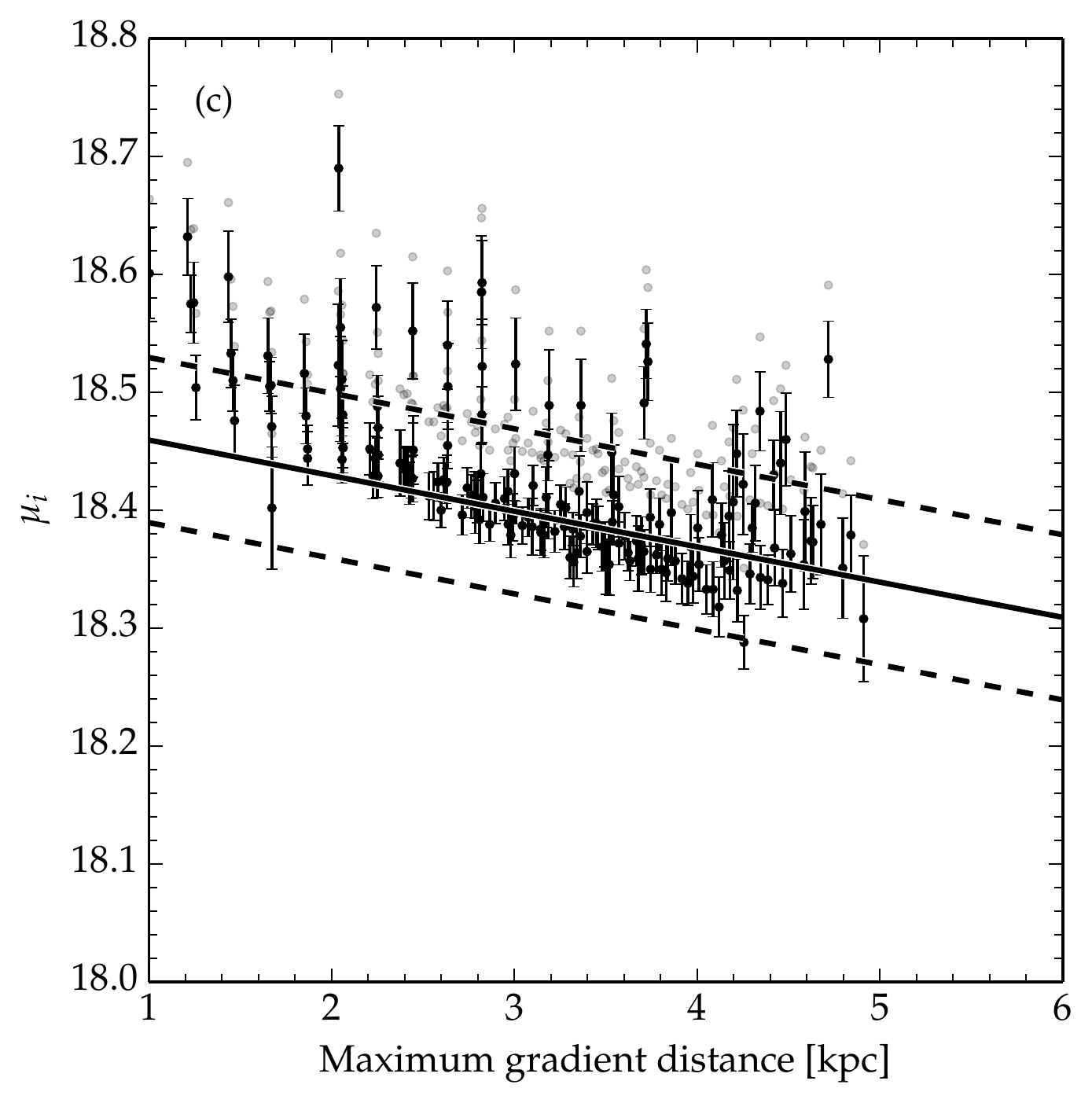}
\includegraphics[width=0.45\textwidth]{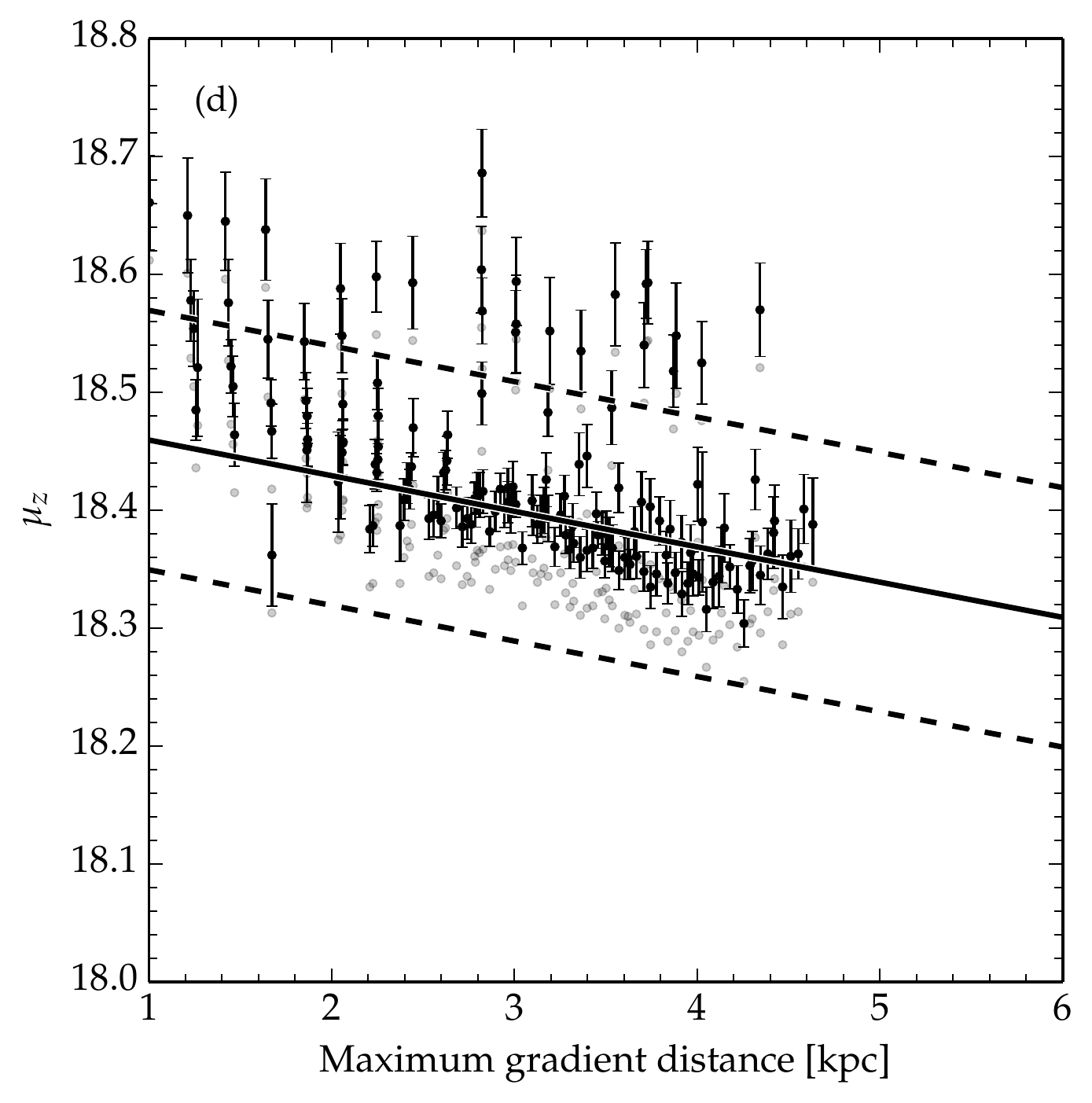}
\caption{The distance modulus as a function of the distance along the maximum
    gradient line ($\sim$ NE-SW). \comm{The black circles are the measured
    distance modulus in each \textsc{HEALpix} pixel, as describe in the text.}
    The error bars shows the 1 $\sigma_m$ for each measured point.  The solid
    line shows the behaviour expected from the best-fit disk model for the
    \emph{all} population. The dashed line shows the maximum magnitude variation
    expected from populations mixing effects. The gray points are the distance
    modulus inferred using the theoretical RC absolute magnitude. Panels a,b,c,
    and d show the distance modulus determination using the filters $g,r,i,$ and $z$
    respectively.}
\label{fig:RC_dmax}
\end{figure*}

To infer the absolute magnitude of the RC $(M_{\lambda}^{RC})$ we could adopt
the prediction from synthetic stellar populations based on stellar evolution
models. However, simulations using \textsc{PARSEC} \citep{Bressan12}
models agree well with the disk models found in this work. A small
magnitude offset can be seen on Figure \ref{fig:RC_dmax}, where the solid
black line shows the best-fit disk model and the gray points are the inferred
distance modulus based on theoretical values of $M_{\lambda}^{RC}$.

Despite the small offset in the theoretical value of $M_{\lambda}^{RC}$ we chose
to determine $M_{\lambda}^{RC}$ by matching the expected distance modulus from
our best-fit disk model to the observed values of the RC peak magnitude.  This
matching was done using stars that are between 3 and 4 kpc along the LMC line of
maximum distance gradient.  We compute the median value of $m_{\lambda}$ in this
distance range and subtract from that the heliocentric distance expected from
the disk model at 3.5 kpc from the LMC centre. This offset is determined for all
passbands. The value obtained is then used to compute the RC-based distances
consistently with our disk model.  These distances are shown as the black points
in Figure \ref{fig:RC_dmax} and the error bars are propagated from the RC
peak fit.


\begin{figure*}
\includegraphics[width=0.45\textwidth]{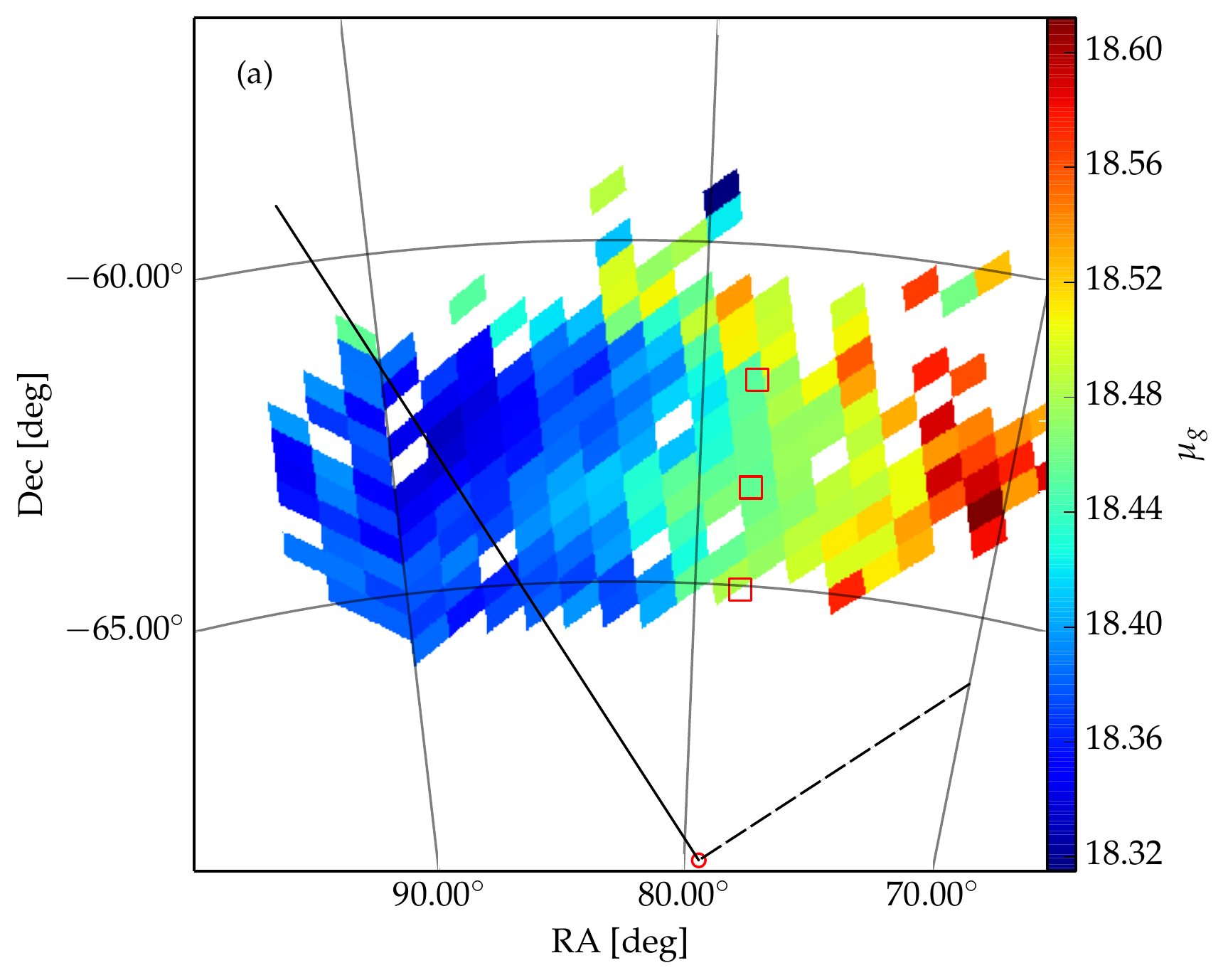}
\includegraphics[width=0.45\textwidth]{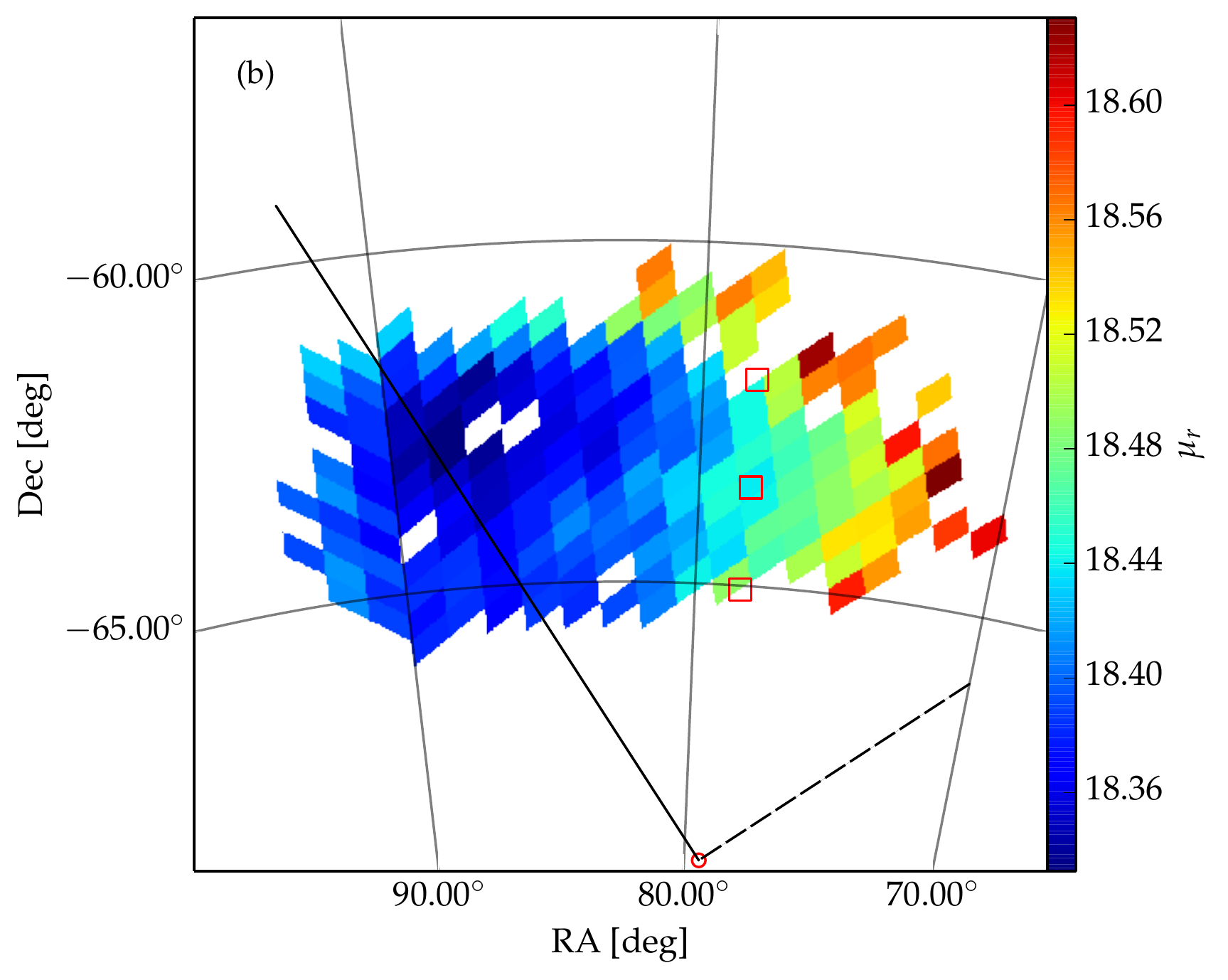} \\
\includegraphics[width=0.45\textwidth]{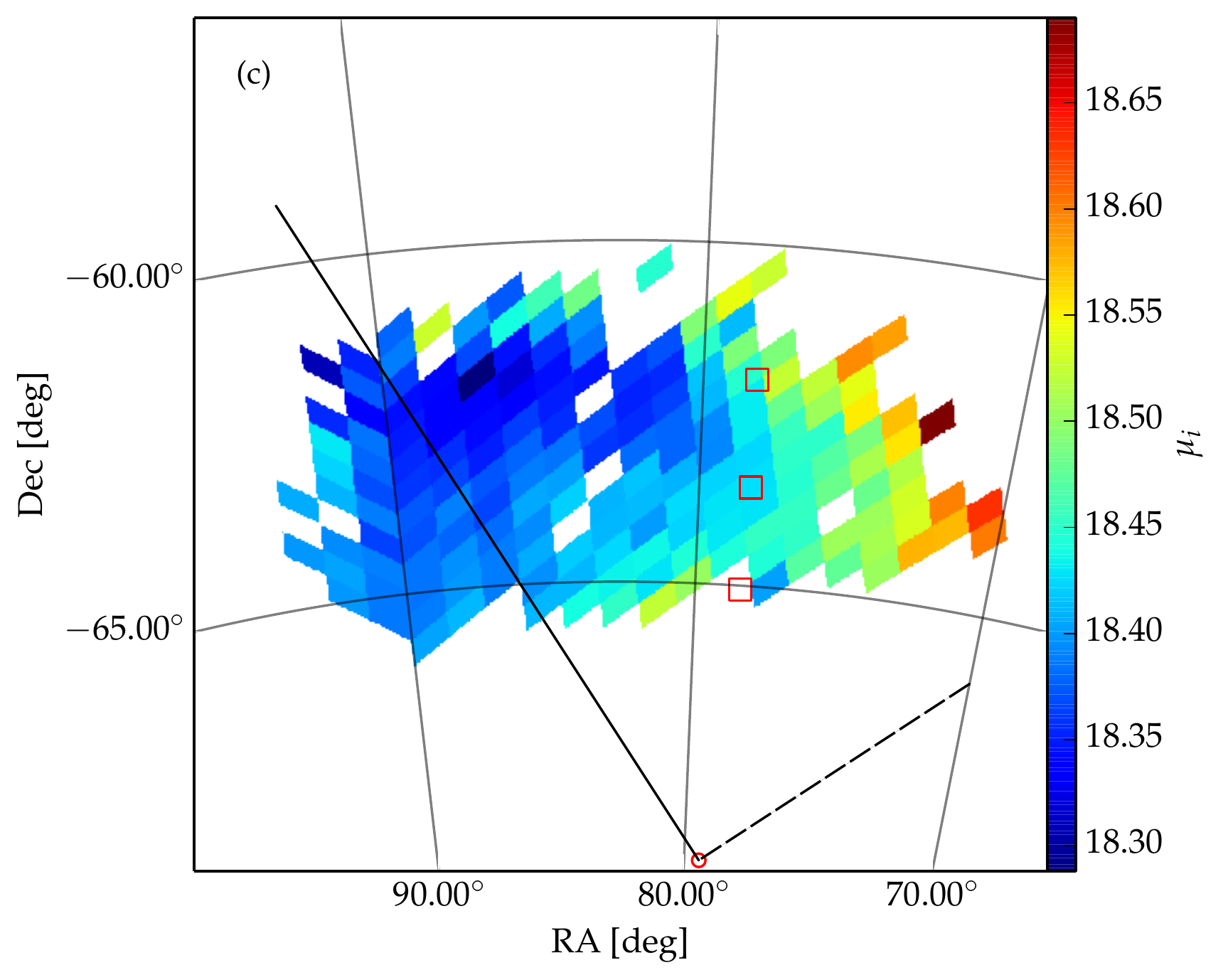}
\includegraphics[width=0.45\textwidth]{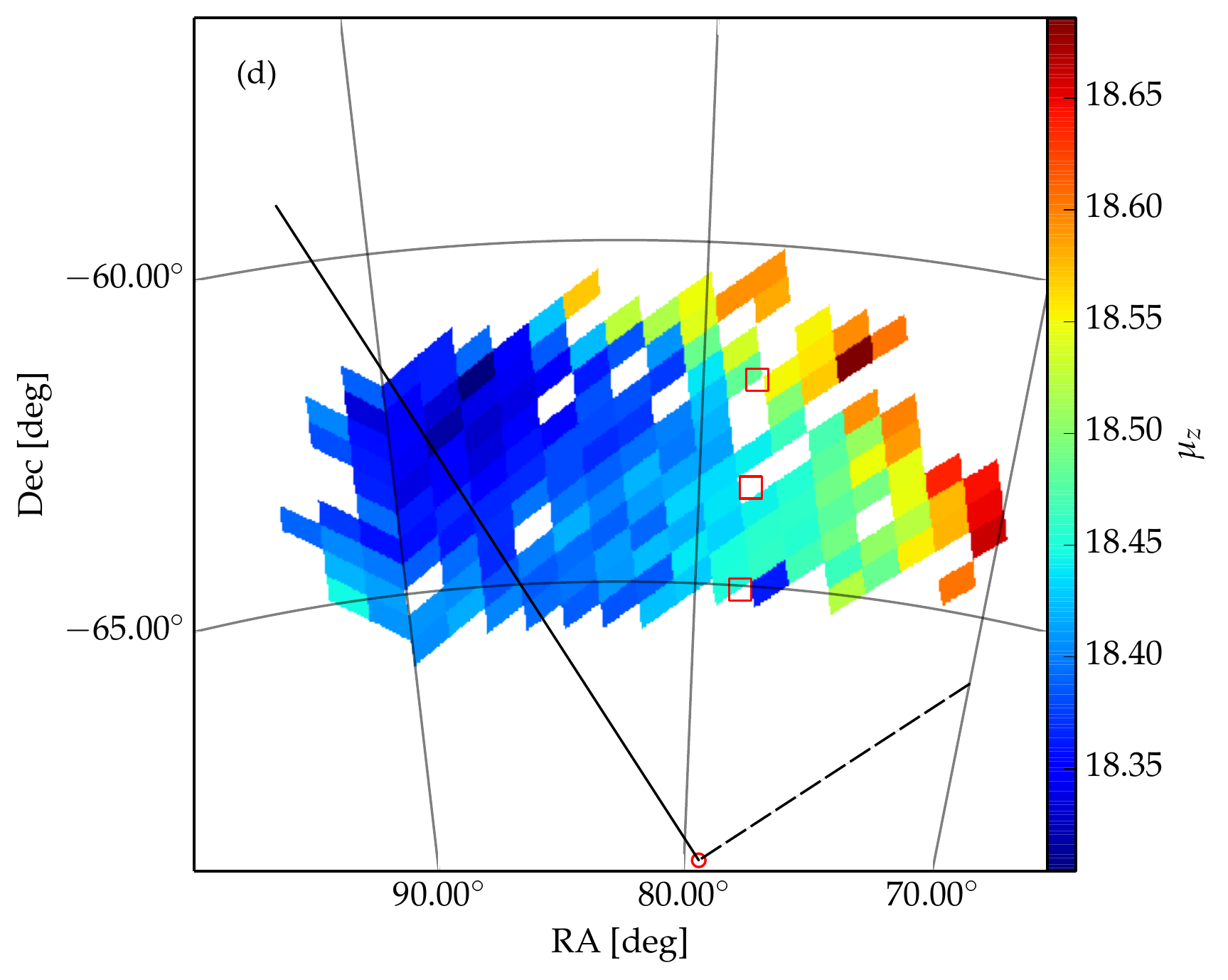}
\caption{Gnomonic projection of $N_{side} = 128$ maps showing the distance
    modulus measured in each the $g,r,i,$and $z$ passband. On every panel the
    solid line shows the direction of the maximum distance gradient
    ($\theta$) and the dashed line the direction of the line of nodes
    ($\theta_0$) expected from the best-fit disk model for the \emph{all}
    population. The red squares are the fields studied by \citet{Meschin14}. The
    red circle is the LMC centre adopted in this work. Panels a, b, c, and d
    show the distance modulus determination using the filters $g$, $r$, $i$, and
    $z$ respectively.}
\label{fig:RC_map}
\end{figure*}

In Figure \ref{fig:RC_map} we show the distributions of $m_{\lambda}^{RC}$
on the sky, for $\lambda = g,\,r,\,i,\,z$. The lines show the direction of
maximum gradient and the line of nodes obtained from our best disk model fit
using the \emph{all} stellar population. We notice that the entire sampled
region in this work is on the same side (near side) of the LMC disk major axis.
There is a global trend in the sense that for all passbands the nearest points
to us are located in the North-East edge of the LMC. Another remarkable feature
present in all maps is that the North edge of the LMC is systematically more
distant than what is expected for the disk model. This effect is noticeable in
all passbands.

The discrepancies observed in the maps described above may be due to
population mixing in RC stars. The RC luminosity depends slightly on
the age and metallicity of the stars \citep{Girardi99, GirardiSalaris00}. This
fact makes the RC of a stellar population older than $\sim 3$ Gyr dimmer as a
function of age. The exact amount of dimming depends on the metallicity and on
the passband used. A detailed discussion about how the RC magnitude of simple
stellar populations (SSP) varies with age and metallicity may be found in
\citet{GirardiSalaris00}.

In the case of the LMC we would like to know how the RC magnitude changes as a
function of the SFH, therefore mapping $\Delta M_{\lambda}^{RC}$ at different
positions in the LMC. However, it is known that the LMC SFH is very complex and
varies spatially \citep{Meschin14}, therefore rendering this a very difficult
task.

To assess the amplitude of the populations mixing effect over the RC
magnitudes, we conducted a series of synthetic stellar population experiments.
These simulations assume that the LMC follows an age-metallicity relation given
by \citet{Piatti12} with a spread of 0.15 dex in $[Fe/H]$. We also assume that
stars in the LMC follow a Kroupa Initial Mass Function (IMF) \citep{Kroupa01}.
To generate synthetic magnitudes for a set of simulated stars we adopt the
\textsc{PARSEC} \citep{Bressan12} isochrones in the DES photometric system. Our
grid of models has a step of 0.01 in $\log_{10}(age/yr)$ and 0.0002 in $Z$ in
the range $[0.0002 ,0.001]$ and 0.001 for Z  in the range $[0.001, 0.020]$.
Photometric uncertainties were included according to Equation
\ref{eq:photerrors} and Table \ref{tab:photerrors}. The simulations were made
in the framework of the open-source code
\textsc{genCMD}\footnote{\texttt{https://github.com/balbinot/genCMD}}

We adopt three SFHs that are modelled according to what has been found by
\citet{Meschin14} for the outer regions of the LMC (red squares in Figure
\ref{fig:RC_map}). The simulated SFHs contain two events of star formation that
were modelled as Gaussian peaks in SFR. The young peak is centered on an age of
2.5 Gyr with a width of 1 Gyr while the second is centered on 9 Gyr with a width
of 1.5 Gyr. The first adopted SFH gives equal amplitude to each star forming
event, the second gives twice the amplitude to the older event, and the third
gives twice the amplitude to the younger event. The largest difference for the
RC peak magnitude was observed between the models with asymmetric peaks in the
SFH. We use this difference to estimate the maximum value for $\Delta
M_{\lambda}^{RC}$ which was found to be 0.06, 0.06, 0.07, and 0.11 for
$g,\,r,\,i,\,z$ respectively. The dashed lines in Figure \ref{fig:RC_dmax} show
the maximum expected deviations due to populations mixing effects.

We notice that the RC distance variations are very similar to the one expected
from the best-fit disk model. However there are very discrepant points that the
disk model cannot account for, even when population mixing effects are
considered. We also notice that the distance moduli determined using different
passbands are consistent.

\begin{figure*}
\includegraphics[width=0.45\textwidth]{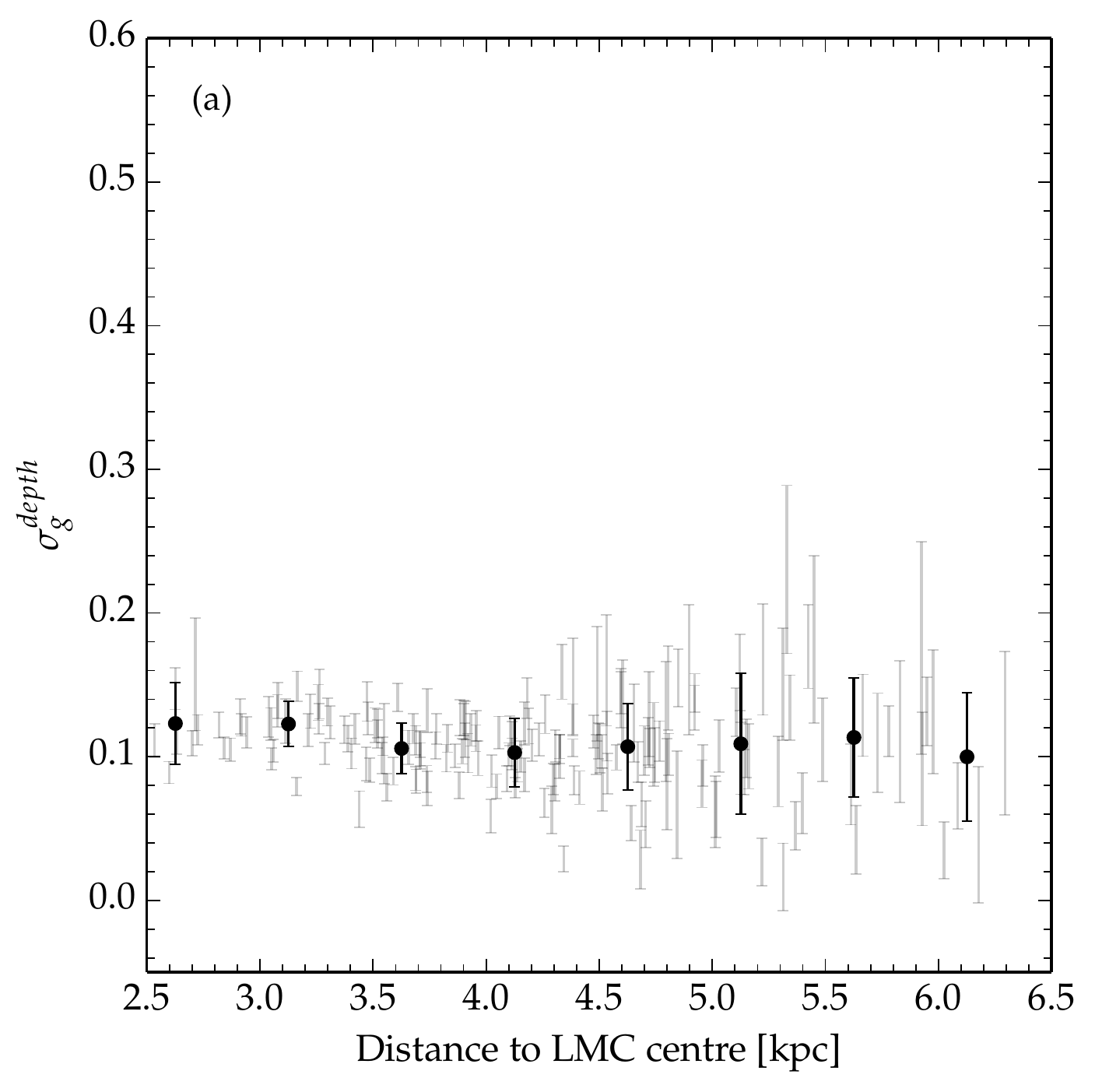}
\includegraphics[width=0.45\textwidth]{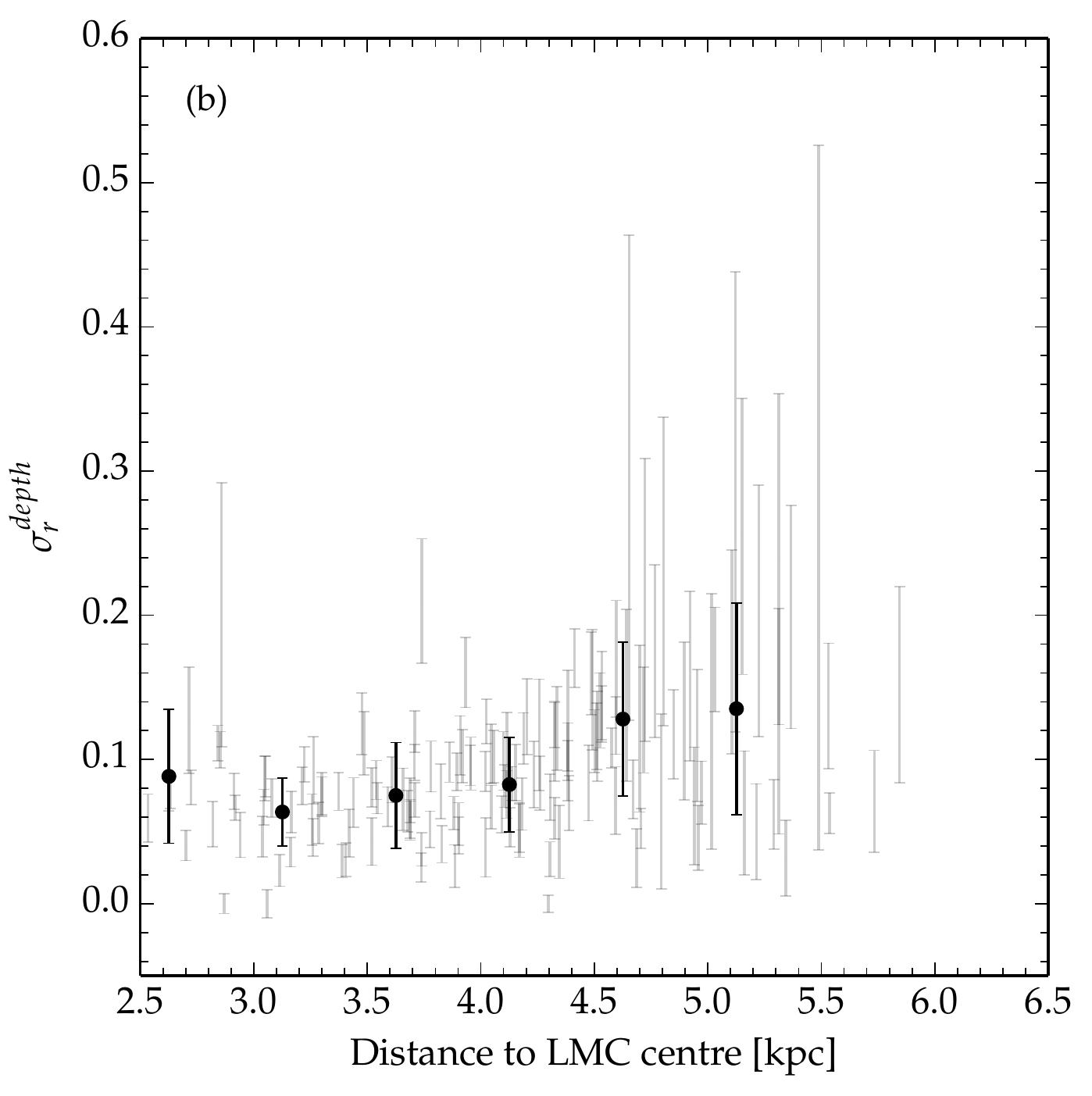} \\
\includegraphics[width=0.45\textwidth]{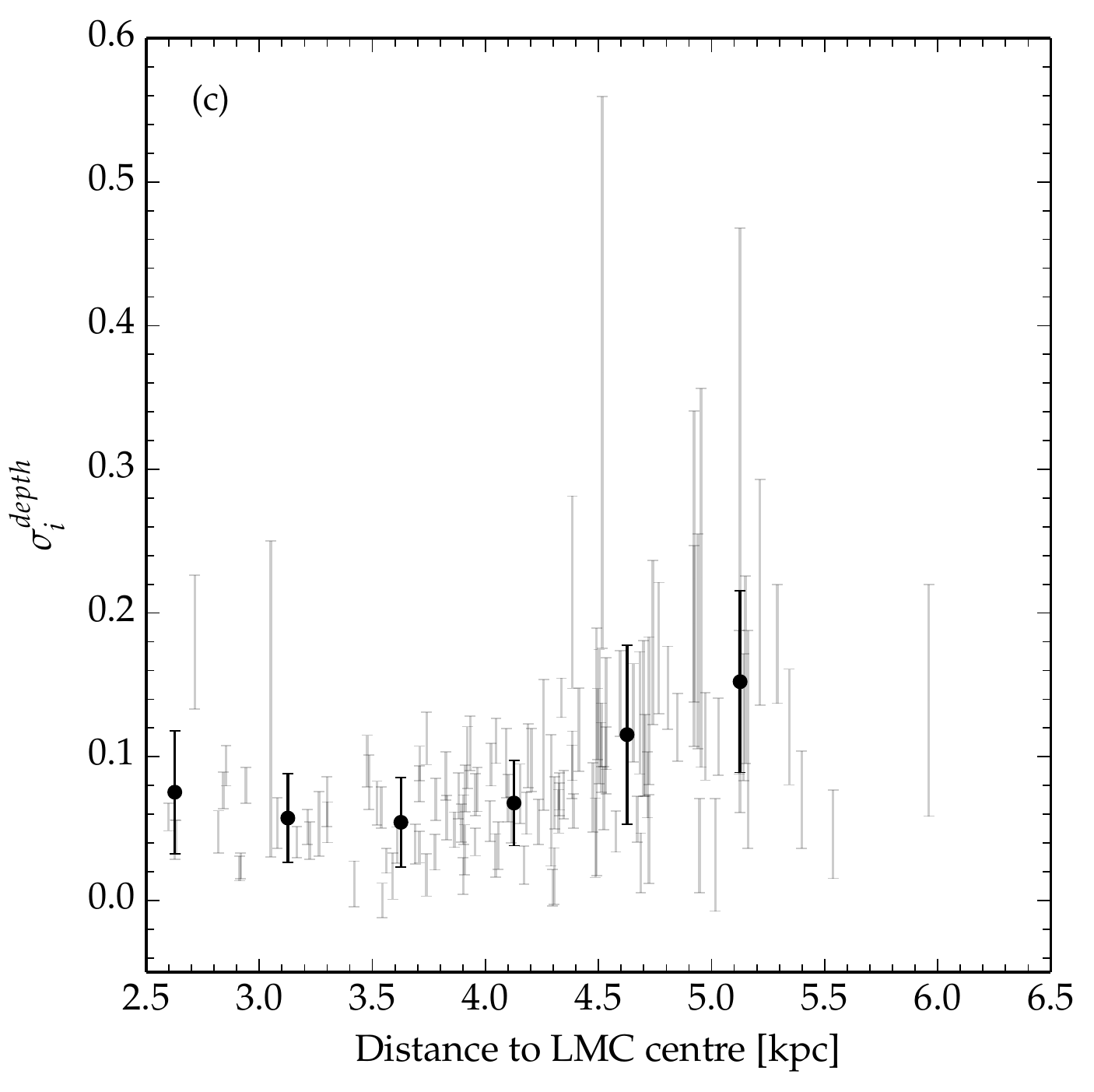}
\includegraphics[width=0.45\textwidth]{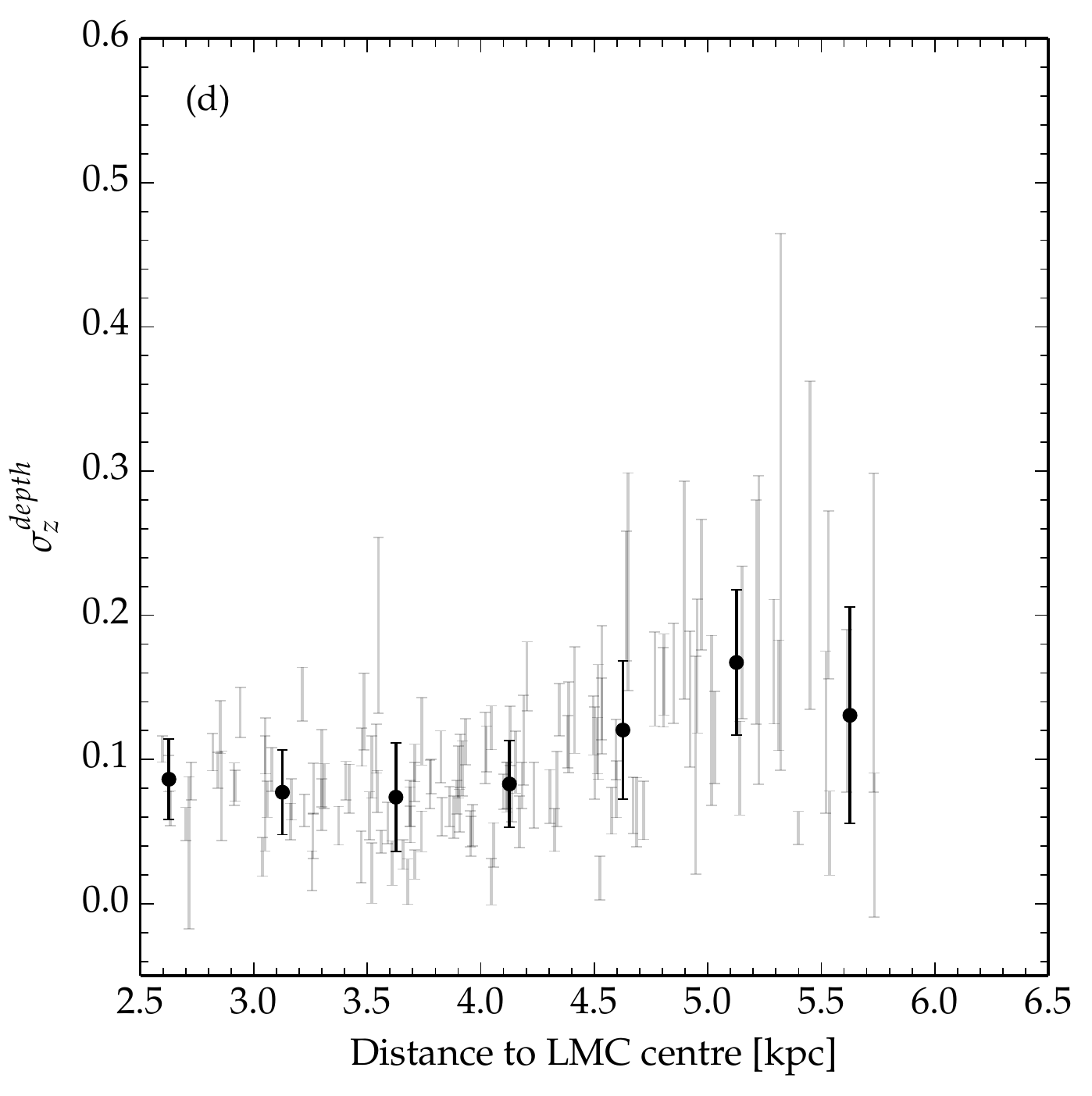}
\caption{Disk depth as a function of distance to the LMC centre. The gray
    error bars show the $\sigma_{depth}$ measured for each \textsc{HEALpix} pixel
    shown on Figure \ref{fig:RC_map}. The dark circles are the average
    over 0.5 kpc and the error bar the standard deviation. Panels a,b,c, and d
    are for determinations using the filters $g,r,i,$ and $z$ respectively.}
\label{fig:RC_depth}
\end{figure*}

From the RC distributions we may also extract information about the thickness of
the LMC disk. This information is embedded in $\sigma_{\lambda}$ shown in
Equation \ref{eq:RC_fit}. However, this quantity is also affected by the
intrinsic scatter of the RC and by photometric uncertainty. The RC simulations
described above incorporate this intrinsic scatter ($\sigma_i$) convolved with
the photometric one ($\sigma_{phot}$). In our simulations we measure this
quantity ($\sqrt{\sigma_i^2 + \sigma_{phot}^2}$) and we find that this quantity
has a mean value of 0.11, 0.12, 0.13, 0.13 for $g,r,i,$ and $z$ respectively.
Finally the last contributions comes from the depth of the disk along the
line of sight itself ($\sigma_{depth}$). Thus, we may write:

\begin{equation}
    \label{eq:sig_lamb}
    \sigma_{depth}^2 = \sigma_{\lambda}^2 - \sigma_{phot}^2 - \sigma_{i}^2.
\end{equation}

In Figure \ref{fig:RC_depth} we show $\sigma_{depth}$ as a function of the
distance along the LMC disk. The error bars are taken from the covariance matrix
obtained on the least-square fit of the RC distribution as given by Equation
\ref{eq:RC_fit}. We took averages of $\sigma_{depth}$ in bins of 0.5 kpc, which
are shown as the black circles. The error bar is the standard deviation of the
values of $\sigma_{depth}$ in each distance bin. For distances less than
$\sim4.5$ kpc we observe a constant depth of $\sim$ 0.08 mag (1.8 kpc) measured
in $r$, $i$, and $z$ while the $g$ band yields a value of 0.12 mag (2.8 kpc).
The depth in $r$, $i$, and $z$ passbands tends to increase by $\sim 0.02$ mag
($\sim 0.5$ kpc) towards the edge of the LMC. The slight difference in the $g$
band depth might be due to the underestimate of the intrinsic RC magnitude
spread. This effect may be due to changes in the metallicity of the LMC field
populations torwards its edge \citep{Majewski09}, leading to the formation of a
Horizontal Branch (HB) instead of a RC. The $g$ band is more affected since it
is more sensitive to hotter stars.

\section{Discussion and summary}

To the best of our knowledge, this paper is the first to show a clear
distinction in the structure of the LMC disk for stellar populations of
different ages. \comm{\citet{Meschin14} report signs of an age gradient,
    but did not derive the structural properties for different age components of
the LMC.} Previous studies, based on large spectroscopic samples, reported a
clear kinematic distinction between different stellar types and attributed that
to an age difference in their samples \citep[][and references
therein]{vdMarel09}. Here we confirm this distinction based only on stellar
photometry alone.

The disk models obtained in this work are in agreement with what has been
reported in the literature \citep[e.g.][]{Rubele12, Saha10, Nikolaev04,
WeinbergNikolaev01}. We observe a significant difference between disk models
fitted using stars younger and older than $4$ Gyr. The most striking
difference is found in the scale length ($R_s$), which suggests a star
formation that follows the outside-in paradigm. The summary
of the disk models parameters can be found in Table \ref{tab:best_fit}.

The \emph{young} population is poorly described by an exponential disk
model. It presents a prominent feature that strongly deviates from an
exponential profile truncated at $\sim$8 kpc. The \emph{young} and \emph{old}
LMC disk have a clear distinction in the level of substructures. The younger
stars are known to have peculiar spiral arms that have most likely been
formed in the last SMC-LMC encounter \citep{Staveley-Smith03,Olsen07,Bekki09}.
This could explain why the \emph{young} disk cannot be well described by our
disk model.

On the other hand, the \emph{old} disk extends over $\sim 13$ scale lengths.
Our results for the scale length and truncation radius are similar to what
\citet{Saha10} found. The authors report $R_s \simeq 1.15$ and a truncation
radius that is $\sim$12 times this value, which is very similar to our
determination. If we assume that the truncation radius of the \emph{old} disk is
caused by the tidal field of the MW, we obtain a LMC mass that is $M_{LMC} =
2.3^{+0.8}_{-0.6} \times 10^{10} M_{\odot}$. This mass is strongly dependent on
the efficiency with which the MW potential tidally truncates the LMC stellar
distribution. Even if the Clouds are in their first pericentric passage, their
orbit is predicted to be past the pericenter, which could allow the formation of
a tidal limit (Besla, private communication). \comm{Also, \citep{Gan10} showed
    that subhalos in their first pericentric passage may undergo substantial
    mass loss. We also point out that in the case where the MW tidal field
    efficiently strips the dark matter and luminous components, our estimate of
    the LMC mass is reliable. However, in the case where dark matter is still
    present in an extended halo, our estimate of the LMC mass sets only a lower
    limit. The absolute mass value is also dependent on the MW mass profile, but
    the relative LMC to MW encircled mass is not. New more precise
    determinations of the MW mass in its outer regions will help to more
accurately constrain the LMC mass.}

The dynamical mass determination by \citet{vdMarel14} suggests $M_{LMC} (R < 8.7
$ kpc$) = (1.7\pm0.7) \times 10^{10} M_{\odot}$ that can be extrapolated to the
mass inside our truncation radius of $M_{LMC} (R < R_t) = (3.5 \pm 1.4) \times
10^{10} M_{\odot}$ assuming a flat rotation curve. This value is within the
error bars of our mass determination, suggesting that the truncation radius
found in this work has a tidal origin.

Through the fit of the RC peak magnitude we observe that a few regions
coherently grouped in the North of the LMC are systematically more distant than
what is expected by an inclined circular disk model. While \citet{Olsen02} found
that regions of the South-East LMC are systematically closer, we find that the
Northern LMC disk behaves in the opposite sense. This is the classic case for a
galaxy with warped disk. We demonstrate, through the use of synthetic stellar
population models, that the offsets from the disk model are larger than what is
expected from population mixing effects.

Using the same synthetic stellar populations quoted above we are able to compute
the theoretical magnitude scatter of RC stars. This allows us to disentangle the
contribution of the intrinsic scatter from the one caused by the disk thickness.
We measure a thickness ranging from 1.8 to 2.8 kpc in the
inner parts of the LMC and which increases by 0.5 kpc in its outer parts. If we
assume that the disk follows an exponential profile in height we obtain that
the scale height ranges from 1.3 to 1.9 kpc, increasing by 0.3 kpc in the outer
parts. This slight increase towards the outskirts can be interpreted as the
\emph{flaring} of the disk. \citet{vdMarel01} argue that this kind of behaviour
is expected if the LMC disk is tidally perturbed by the MW potential. The disk
thickening may also be explained solely by LMC dynamical models
\citep{Besla12}, where the old stellar component is expected to form a
dynamically hotter component. The fact that we only observe it on the edge of
the LMC may be due to the steep truncation of the \emph{young} disk, allowing
us to measure a thickness that is dominated by the \emph{old} population only
in the LMC outskirts.

The analysis presented in this work points to a scenario where there are two
distinct disk components in the LMC. One is composed of old stars ($>
4$ Gyr) and possesses a smooth, extended profile out to dozens of scale
lengths. The second component is composed of younger stars ($< 4$ Gyr) and
appears to be pertubed and much less extended. We find signs of \emph{warp}
and \emph{flare} towards the outskirts of the LMC, both of which have been
reported previously \citep{Olsen02, Subramaniam09}.

An alternative scenario for the disk \emph{warp} and \emph{flare} is the
existence of a hot halo composed of \emph{old} stars, while the disk could
contain both populations. Because the \emph{old} population is more
\emph{spatially} extended, this scenario would explain the outward increase in
the thickness of the RC as the \emph{young} disk fades, therefore accounting
for the ``flare'' signal.  Besides the {\it flaring}, the \emph{old} spheroidal
component would lead to a systematic variation on the heliocentric distances
across the LMC in the sense of bringing them closer to the heliocentric
distance of the LMC centre on both sides of the galaxy, also mimicking a
\emph{warp} effect.

The fact that the \emph{old} disk is well-fit by a single exponential disk
favours the scenario where there is no detectable stellar halo and older stars
form a thicker disk. Also the fact that the hot halo has only been possibly
detected in the central regions of the LMC \citep{Minniti03} raises doubts about
the existence of such a stellar component. To investigate the existence of a
halo component, a more detailed study must be carried out in the LMC outskirts
in order to isolate its extremely sparse stellar population. This kind of study
should employ more sophisticated CMD decontamination methods such as the ones
used to detect streams and tidal tails \citep{Rockosi02}.

The upcoming years promise to greatly increase our understanding of the
Magellanic system. With the data already observed during the DES SV campaign,
we have the ability to study a vast number of LMC star clusters and their
3-dimensional distribution in the galaxy. The field LMC population may also be
used to infer a very detailed, and spatially dependent SFH.

Upcoming DES observations will cover a large portion of the Magellanic
Stream allowing for the discovery and characterization of its stellar
component, if it exists. Additionally, DES will continue to map the outskirts
of the LMC and SMC. This novel dataset will certainly reveal some of the many
puzzles of the Magellanic system.

\section*{Acknowledgments}

This paper has gone through internal review by the DES collaboration.

We are grateful for the extraordinary contributions of our CTIO colleagues and
the DECam, Commissioning and Science Verification teams in achieving the
excellent instrument and telescope conditions that have made this work possible.
The success of this project also relies critically on the expertise and
dedication of the DES Data Management group.

Funding for the DES Projects has been provided by the U.S. Department of
Energy, the U.S. National Science Foundation, the Ministry of Science and
Education of Spain, the Science and Technology Facilities Council of the United
Kingdom, the Higher Education Funding Council for England, the National Center
for Supercomputing Applications at the University of Illinois at
Urbana-Champaign, the Kavli Institute of Cosmological Physics at the University
of Chicago, Financiadora de Estudos e Projetos, Funda{\c c}{\~a}o Carlos Chagas
Filho de Amparo {\`a} Pesquisa do Estado do Rio de Janeiro, Conselho Nacional
de Desenvolvimento Cient{\'i}fico e Tecnol{\'o}gico and the Minist{\'e}rio da
Ci{\^e}ncia e Tecnologia, the Deutsche Forschungsgemeinschaft and the
Collaborating Institutions in the Dark Energy Survey.

The Collaborating Institutions are Argonne National Laboratory, the University
of California at Santa Cruz, the University of Cambridge, Centro de
Investigaciones Energeticas, Medioambientales y Tecnologicas-Madrid, the
University of Chicago, University College London, the DES-Brazil Consortium,
the Eidgen{\"o}ssische Technische Hochschule (ETH) Z{\"u}rich, Fermi National
Accelerator Laboratory, the University of Edinburgh, the University of Illinois
at Urbana-Champaign, the Institut de Ciencies de l'Espai (IEEC/CSIC), the
Institut de Fisica d'Altes Energies, Lawrence Berkeley National Laboratory, the
Ludwig-Maximilians Universit{\"a}t and the associated Excellence Cluster
Universe, the University of Michigan, the National Optical Astronomy
Observatory, the University of Nottingham, The Ohio State University, the
University of Pennsylvania, the University of Portsmouth, SLAC National
Accelerator Laboratory, Stanford University, the University of Sussex, and
Texas A\&M University.

The DES participants from Spanish institutions are partially supported by
MINECO under grants AYA2009-13936, AYA2012-39559, AYA2012-39620, and
FPA2012-39684, which include FEDER funds from the European Union.

DG was supported by SFB-Transregio 33 'The Dark Universe' by the Deutsche
Forschungsgemeinschaft (DFG) and the DFG cluster of excellence 'Origin and
Structure of the Universe'.

JZ acknowledges support from the European Research Council in the form of a
Starting Grant with number 240672.

\bibliographystyle{mn2e/mn2e}       
\bibliography{refs}                 

\section*{Affiliation}

$^1$Department of Physics, University of Surrey, Guildford GU2 7XH, UK \\
$^2$Departamento de Astronomia, Universidade Federal do Rio Grande do Sul, Av.
Bento Gon\c{c}alves 9500, Porto Alegre 91501-970, RS, Brazil \\
$^3$Laborat\'orio Interinstitucional de e-Astronomia - LIneA, Rua Gal. Jos\'e
Cristino 77, Rio de Janeiro, RJ - 20921-400, Brazil \\
$^4$Osservatorio Astronomico di Padova – INAF, Vicolo dell'Osservatorio 5,
I-35122 Padova, Italy \\
$^5$Observat\'orio Nacional, Rua Gal. Jos\'e Cristino 77, Rio de Janeiro, RJ -
20921-400, Brazil \\
$^6$Astronomy Department, University of Illinois, 1002 W. Green St., Urbana, IL
61801, USA \\
$^7$National Center for Supercomputing Applications, University of Illinois, 1205 W
Clark St., Urbana, IL 61801, USA \\
$^8$Cerro Tololo Inter-American Observatory, National Optical Astronomy
Observatory, Casilla 603, La Serena, Chile \\
$^9$Fermi National Accelerator Laboratory, P.O. Box 500, Batavia, IL 60510 USA \\
$^{10}$Department of Physics \& Astronomy, University College London, Gower Street,
London WC1E 6BT, UK \\
$^{11}$Space Telescope Science Institute (STScI), 3700 San Martin Drive,
Baltimore, MD  21218 \\
$^{12}$Argonne National Laboratory, 9700 South Cass Avenue, Lemont, IL 60439, USA \\
$^{13}$Carnegie Observatories, 813 Santa Barbara St., Pasadena, CA 91101, USA \\
$^{14}$Institut d'Astrophysique de Paris, Univ. Pierre et Marie Curie \& CNRS 
UMR7095, F-75014 Paris, France \\
$^{15}$Kavli Institute for Particle Astrophysics and Cosmology 452 Lomita Mall,
Stanford University, Stanford, CA, 94305 \\
$^{16}$George P. and Cynthia Woods Mitchell Institute for Fundamental Physics
and Astronomy, and Department of Physics and Astronomy, Texas A\&M University,
College Station, TX 77843, USA \\
$^{17}$Department of Physics, Ludwig-Maximilians-Universit\"at, Scheinerstr. 1,
81679 M\"unchen, Germany \\
$^{18}$Excellence Cluster Universe, Boltzmannstr. 2, 85748 Garching, Germany \\
$^{19}$Department of Physics, University of Michigan, Ann Arbor, MI 48109, USA \\
$^{20}$Department of Astronomy, University of Michigan, Ann Arbor, MI 48109,
USA \\
$^{21}$Max Planck Institute for Extraterrestrial Physics, Giessenbachstrasse,
85748 Garching, Germany \\
$^{22}$Department of Physics, The Ohio State University, Columbus, OH 43210,
USA \\
$^{23}$Australian Astronomical Observatory, North Ryde, NSW 2113, Australia \\
$^{24}$Department of Physics and Astronomy, Center for Particle Cosmology,
University of Pennsylvania, 209 South 33rd Street, Philadelphia, PA 19104, USA \\
$^{25}$Institut de F\'isica d'Altes Energies, Universitat Aut\`onoma de
Barcelona, E-08193 Bellaterra (Barcelona), Spain \\
$^{26}$ Instituci\'o Catalana de Recerca i Estudis Avan\c{c}ats, E-08010
Barcelona, Spain \\
$^{27}$Brookhaven National Laboratory, Bldg 510, Upton, NY 11973, USA \\
$^{28}$Centro de Investigaciones Energ\'eticas, Medioambientales y
Tecnol\'ogicas (CIEMAT), Av. Complutense 40, 28040 Madrid, Spain \\
$^{29}$SLAC National Accelerator Laboratory, Menlo Park, CA 94025, USA \\
$^{30}$Jodrell Bank Centre for Astrophysics, University of Manchester, Alan
Turing Building, Manchester, M13 9PL, U.K. \\

\label{lastpage}
\end{document}